\documentclass[preprint]{aastex62}
\usepackage{natbib}
\usepackage{mathtools}
\usepackage{color}
\shorttitle{Powerful BAL outflow in SDSS~J1352+4239}
\shortauthors{Choi et al.}

\begin{document}


\title{Discovery of a Remarkably Powerful Broad Absorption Line Quasar Outflow in SDSS~J135246.37+423923.5}


\author{Hyunseop\ Choi}
\affiliation{Homer L.\ Dodge Department of Physics and Astronomy, The
  University of Oklahoma, 440 W.\ Brooks St., Norman, OK 73019}
\author{Karen M.\ Leighly}
\affiliation{Homer L.\ Dodge Department of Physics and Astronomy, The
  University of Oklahoma, 440 W.\ Brooks St., Norman, OK 73019}
\author{Donald M.\ Terndrup}
\affiliation{Homer L.\ Dodge Department of Physics and Astronomy, The
  University of Oklahoma, 440 W.\ Brooks St., Norman, OK 73019}
\affiliation{Department of Astronomy, The Ohio State University, 140
  W.\ 18th Ave., Columbus, OH 43210}
\author{Sarah C.\ Gallagher}
\affiliation{Department of Physics \& Astronomy, The University of Western
  Ontario, London, ON, N6A 3K7, Canada}
\affiliation{Canadian Space Agency, 6767 Route de l'Aeroport,
  Saint-Hubert, Quebec, J3Y~BY9}
\affiliation{Institute for Earth and Space Exploration, The
  University of Western Ontario, London, ON, N6A 3K7, Canada}
\affiliation{The Rotman Institute of Philosophy, The University of
  Western Ontario, London, ON, N6A 3K7, Canada}
\author{Gordon T.\ Richards}
\affiliation{Department of Physics, Drexel University, 32 S.\ 32nd St.,
  Philadelphia, PA 19104}  



\begin{abstract}
Broad absorption line (BAL) features in quasar spectra reveal an unambiguous signature of energetic outflows from central supermassive black holes, and thus BAL quasars are prime targets for investigating the potential process of luminous quasar feedback on galaxies.
We analyzed the rest-UV spectrum of an ``overlapping trough'' iron low-ionization broad absorption line quasar (FeLoBAL) SDSS~J135246.37+423923.5 using the novel spectral synthesis code {\it SimBAL} \citep{leighly18a} and discovered an extraordinarily fast and energetic BAL outflow.
Our analysis revealed outflow velocities reaching $\sim -38000\rm \, km\, s^{-1}$ with a velocity width of $\sim 10000\rm \, km\, s^{-1}$ which is the largest FeLoBAL outflow velocity measured to date.
The column density of the outflow gas is log$N_H\sim23.2\,[\rm cm^{-1}]$ with the log kinetic luminosity $\log L_{KE}\sim48.1$ [erg $\rm s^{-1}$] which exceeds the bolometric luminosity of the quasar and is energetic enough to effectively drive quasar feedback.
The energy estimate for the outflow is far greater than the estimates from any BAL object previously reported.

The object also shows ``anomalous reddening'' and a significant scattered component that we were able to model with {\it SimBAL}.
We found the first definitive case for radiation filtering in an additional zero-velocity absorption component that required an absorbed continuum to produce the particular absorption lines observed (\ion{Mg}{2}, \ion{Al}{3} and \ion{Al}{2}) without also producing the high ionization lines such as \ion{C}{4}.

\end{abstract}

\section{Introduction}\label{intro}
Broad absorption line (BAL) quasars (BALQs) have been studied extensively in the past several decades since their discovery \citep{lynds67}, and their distinctive blueshifted BAL features provide clear evidence for quasar outflows \citep[e.g.,][]{weymann91}.
Once corrected for selection effects, BALQs are found in 20\%$\sim$40\% of the total quasar population \citep{foltz90,weymann91,tolea02,reichard03,trump06,dai08,knigge08,allen11}.
BALQs are further divided into subgroups based on their spectroscopic properties.
High-ionization BALQs (HiBALs) show only the absorption transitions from highly ionized atoms (\ion{C}{4}, \ion{Si}{4}, \ion{N}{5}, \ion{O}{6}), while low-ionization (LoBALQs) show both the high-ionization transitions and absorption lines from lower-ionization ions (\ion{Mg}{2}, \ion{Al}{2}, \ion{Al}{3}) in their rest-UV spectra.
There is also another class of rarer BALQs called FeLoBALQs that show \ion{Fe}{2} absorption lines.
These objects have large gas column densities, thick enough to extend beyond the hydrogen ionization front \citep{hazard87}.
Although FeLoBALs comprise less than $\sim$ 2\% of the observed quasar population \citep{dai12}, their outflows can have the highest column densities compared to other types of BAL outflows \citep{lucy14}.
Some FeLoBAL objects with broad saturated troughs, where the troughs overlap to nearly completely absorb the continuum emission shortward of 2800 \AA\/, are called `overlapping trough' objects \citep[e.g.,][]{hall02}, and they are expected to have the largest hydrogen column densities (log $N_H$) in their outflows.

Outflowing winds with energy exceeding 0.5\%$\sim$5\% of the quasar luminosity \citep[e.g.,][]{scannapieco04,dimatteo05,hopkins10} are thought to be able to effectively cause AGN feedback.
Outflow energies depend on the amount of material ($\log N_H$) that is being carried by the wind, and more importantly, the velocity of the outflow through $\dot E_k=8\pi \mu m_p \Omega R N_H v^3$ \citep{dunn10}.
The combination of large column density ($\log N_H$) and high velocity produce energetic outflows.

A few discoveries of high-velocity HiBAL outflows ($v\sim$ 0.1c--0.3c) have been made.
For example, \citet{rodri09} discussed a $v\sim$ 0.2c BAL outflow in PG0935+417 and \citet{hamann18} suggested that there is a \ion{C}{4} BAL feature at $v\sim$ 0.3c in PDS 456.
\citet{rogerson16} reported BAL features at $v\sim$ 0.2c and 0.1c in the variable HiBALQ SDSS~0230$+$0059.
In the cases mentioned above, the physical properties of the outflows were not sufficiently constrained to estimate the outflow energy because those HiBAL objects only showed prominent \ion{C}{4} absorption lines (and \ion{Si}{4} or \ion{N}{5} lines in some cases) and lacked diagnostic lines to probe the density of the outflow.
Moreover, HiBALQs are not expected to have the highest $\log N_H$.

LoBALQs and FeLoBALQs have significantly higher column densities, and therefore, high-velocity outflows in these objects may yield produce the most energetic outflows.
\citet{borguet13} and \citet{chamberlain15} analyzed the rest-UV spectra of LoBALQs SDSS~J1106+1939 and SDSS~J0831+0354, respectively.
They found high-velocity LoBAL outflows with high energies and constrained their physical properties ($\sim-8000\rm \, km\, s^{-1}$ and $\sim-10000\rm \, km\, s^{-1}$, respectively; see \S~\ref{postpro}).
Although the FeLoBALs are expected to have thick (highest log $N_H$) and massive outflows, potentially harboring energetic outflows, only a few FeLoBAL objects have been analyzed to determine the physical properties of their outflows \citep{dekool01,dekool02f1,dekool02f2,dunn10,bautista10,lucy14}.
Because the common method \citep[e.g.,][]{arav13} used to analyze BAL troughs involves individual line identification, it becomes extremely challenging to extract physical properties of an outflow that has a large number of \ion{Fe}{2} absorption features that are blended together.

{\it SimBAL} was first introduced by \citet{leighly18a} as a novel spectral synthesis code developed to analyze BAL outflows.
Because {\it SimBAL} uses forward modeling with spectral synthesis, the code can be used to analyze even the most complex BAL spectroscopic features with significant line blending.
The code has produced an excellent fit to SDSS~J0850+4451 \citep{leighly18a}, a LoBAL object; moreover its sophisticated treatment of modeling the partial coverage of BAL absorbers led to further understanding of the geometry and the structure of the outflow \citep{leighly18b}.

For thick BAL outflows, part of the radiation can be significantly absorbed by gas closer to the central engine before reaching the gas further away producing a phenomenon called ``radiation filtering or shielding'' \citep[e.g.,][for the case of emission lines]{leighly04,leighly07}.
The question of whether or not the radiation filtering is important in outflows has gained some recent attention.
\citet{leighly18a} recently explored the possibility of radiation filtering in their {\it SimBAL} models and found no evidence supporting the phenomenon in SDSS~J0850+4451.
\citet{miller18} suggested a potential two-phase photoionization condition arising from radiation filtering in LBQS~1206+1052.
Despite the effort to understand the radiation filtering, no definitive observational evidence has been found.

Not only do BALQs show interesting outflow signatures, they also are known to show stronger reddening and a higher scattering fraction \citep[e.g.,][]{sprayberry92,brotherton97,dipompeo11,krawczyk15}.
Some extragalactic objects are known to show ``anomalous reddening'', where their reddening curves do not resemble any of the commonly used reddening curves derived from the Milky Way galaxy \citep[e.g.,][]{cardelli89} or the Magellanic Clouds \citep[e.g.,][]{prevot84}, possibly due to a particular dust composition near the quasar \citep{hall02,leighly09,jiang13,fynbo13,zhang15,krogager15,meusinger16}.
The nature of the strong reddening observed in BALQs may offer clues to the physical conditions and geometry of the outflows in these objects.
Moreover, the dust has significantly larger scattering cross-section than the ions and can provide significant acceleration to the outflows \citep[e.g.,][]{fabian08,fabian18}.
Dusty outflows are able to harness the radiation pressure more efficiently and could potentially explain the acceleration mechanism of some of the BAL outflows with the highest velocities.

In this paper, we report the discovery of the most energetic BAL outflow analyzed to date.
SDSS~J135246.37+423923.5, hereafter referred to as SDSS~1352$+$4239, is an overlapping trough object that was initially observed by the Sloan Digital Sky Survey (SDSS).
This object has all the fascinating BAL characteristics in its spectrum, including a wide overlapping trough, anomalous reddening and a substantial scattered light signature.
With new near-infrared observations of SDSS~1352$+$4239, we measured an accurate redshift, z=2.26, from the Balmer emission lines.
From the correct redshift we were able to identify the fastest FeLoBAL outflow ever observed ($v\ \sim -38000\rm \, km\, s^{-1}$).
We performed detailed analysis with {\it SimBAL} to determine the physical conditions of the outflowing cloud and constrain the energetics of the outflow.
We were able to not only characterize the main BAL outflow but we also found evidence for radiation shielding in the zero-velocity BAL system.
In \S~\ref{simbal}, we briefly reintroduce {\it SimBAL} and the changes that have been made since its debut in \citet{leighly18a}.
In \S~\ref{observations}, we describe the new observation and data reduction done for SDSS~1352$+$4239.
We introduce a general reddening curve used to model the unusual continuum shape in \S~\ref{contsedmod} and we describe the spectral model used with {\it SimBAL} to analyze SDSS~1352$+$4239 in \S~\ref{bestmodel}.
We report the energetics derived from the {\it SimBAL} fit of the outflow in \S~\ref{postpro} and compare our result with other quasar objects known to have powerful outflows.
Implications of our findings and a summary can be found in \S~\ref{disc} and \S~\ref{conclusions}.

\section{{\it SimBAL}}\label{simbal}
Constraining the physical conditions of the outflowing clouds can be very challenging due to line blending and the non-black saturation of absorption lines from partial coverage of the emission sources.
The standard method for analyzing BALQ spectra relies on the apparent optical depth (AOD) analysis \citep[e.g.,][]{arav13}.
This method requires line identification and optical depth measurement of each absorption line.
The optical depths are converted to ionic column densities and compared to the output from 1D photoionizations simulations using {\it Cloudy} \citep{ferland17} to find the physical conditions of the gas along the line of sight.
Because the AOD analysis can only provide lower limits for the column density estimates for the identified absorption lines and fails to provide accurate line ratios due to non-black saturation, accurate measurements of the density and the location of the gas with respect to the ionizing continuum source is difficult.

An alternative approach to studying BALQ spectra with the novel spectral synthesis code {\it SimBAL} was introduced by \citet{leighly18a}.
{\it SimBAL} uses grids of ionic column densities calculated using the photoionization code {\it Cloudy} \citep{ferland17} and a Bayesian model calibration method to model BALQ spectra.
Because {\it SimBAL} employs a forward modeling technique and a sophisticated mathematical implementation of partial covering to model the absorption features \citep{leighly18b}, it can accurately reproduce the complex absorption features in BALQSOs and constrain the physical properties of the outflow as a function of velocity.
With a given set of parameters, {\it SimBAL} combines ionic column density information from the {\it Cloudy} grids, line transition strengths from atomic data and the parameterized kinematics of the outflow to create a synthetic spectrum.
Additionally, the Bayesian model calibration method used in {\it SimBAL} yields error estimates for the physical parameters that describe the gas in the outflow.
A detailed discussion on how {\it SimBAL} operates and a flowchart describing the relationship of the components can be found in \S~3 and Figure~2 of \citet{leighly18a}, and we review the basic features here.

Each absorption component is specified by 6 parameters: ionization parameter log $U$, density $\log n\rm\,[cm^{-3}]$, thickness of the gas relative to the hydrogen ionization front $\log N_H\,-\,\log U\rm\,[cm^{-2}]$, outflow velocity $v\rm \, (km\, s^{-1})$, velocity width $\sigma\rm \, (km\, s^{-1})$, and a covering fraction parameter log $a$ (discussed further below).
The first three parameters define the physical conditions of the outflowing gas in terms of the photoionization state and the last three parameters define the kinematics of the gas as well as the state of non-black saturation by modeling the partial coverage using the covering fraction parameter.
{\it SimBAL} can model a broad absorption feature with either one or multiple Gaussian opacity profiles or the ``tophat accordion'' model where a broad velocity profile is divided up into multiple velocity-adjacent ``tophat'' bins \citep{leighly18a}.
The number of bins is fixed for a given model.
Each bin can have its own set of physical parameters (i.e., ionization parameter, density and log $N_H$ $\sbond$ log $U$) and a covering-fraction parameter.
Alternatively, parameters can be tied together for several velocity bins.
As discussed in detail in \citet{leighly18b}, the inhomogeneous partial covering model in {\it SimBAL} uses a powerlaw distribution of opacity $\tau$ where $\tau =\tau_{max}x^a$ \citep{sabra05,arav05}.
{\it SimBAL} uses log $a$ to control the partial coverage and $x\,\in\,(0,1)$ in the above equation is a normalized continuum source size scale.
Full covering is achieved with low values of $a$ close to 0, and low covering can be modeled with high values of $a$.
Further discussion of inhomogeneous partial covering is given in \citet{leighly18b}.

The version of {\it SimBAL} used in \citet{leighly18a,leighly18b} used the 2013 version of {\it Cloudy}.
After that analysis was initiated, version C17 of {\it Cloudy} \citep{ferland17} was released, which allowed more complete and accurate photoionization calculations with a significantly larger atomic database.
Compared to \citet{leighly18a}, the ionic column density grids that have been calculated with version C17 of {\it Cloudy} include the column densities of \ion{Fe}{2} ions with a greater number of excited state levels and multiple iron-peak element ions including Co and Zn at multiple ionization states.
{\it SimBAL} previously used a line list with 6267 transitions (78 ions; 179 counting the number of excited energy states); the updated line list includes 76488 transitions (281 ions; 997 counting the number of excited energy states).
A second update from the previous version of {\it SimBAL} involves the grid sampling.
The photoionization state of the gas changes dramatically near the hydrogen ionization front.
A simple even sampling by a modest amount across the column density or the $\log N_H\,-\,\log U$ parameter is insufficient to characterize the rapid change of ionic column densities across the hydrogen ionization front.
For example, the ionic column densities of some species that are mostly found in the partially ionized zone such as \ion{Fe}{2} increase by more than 4 dex as the hydrogen ionization front is traversed \citep[e.g.,][their Fig.~10]{lucy14}.
A finer sampling is needed to properly capture the steep increase in ionic column density around the hydrogen ionization front.
However, the remainder of the hydrogen column density range does not need a finer sampling and a grid with much finer sampling requires a tremendous amount of calculation time as well as a large file size.
Therefore we approach this problem by adopting a flexible indexing scheme where we identify the location of the hydrogen ionization front and apply the oversampling only around the region where the ionic column densities change very rapidly.
In addition, the changes in physical condition before and after the hydrogen ionization front becomes more dramatic with higher ionization parameter.
We took into account this change in the ``sharpness'' of the hydrogen ionization front when calculating the indexing scheme by increasing the grid density of the oversampled regions for higher ionization parameters (total 619,721 grid points).

A third change involves continuum modeling of the spectra.
In \citet{leighly18a}, continuum-normalized spectra were used for analysis.
The issue is that the depth of the absorption feature can either be overestimated or underestimated depending on the continuum placement.
The new version of {\it SimBAL} models both the synthetic continuum model and the absorption model simultaneously, producing a full synthetic spectrum to be compared with the data as well as the unabsorbed spectrum model.
Thus {\it SimBAL} can fit both the emission features and the absorption features of the spectrum simultaneously to produce a more robust solution.
This methodology allows more accurate measurement of the outflows.
Moreover, simultaneous absorption and emission continuum modeling enables the fitting of heavily absorbed objects (e.g., overlapping trough objects) that have thick outflows and show very little residual continuum emission.
In this paper, we use an emission line template developed from an HST observation of Mrk 493 (\S~\ref{modelem}).
More generally, we use principal component analysis (PCA) eigenvectors for the continuum modeling with {\it SimBAL} \citep{leighly_prep}.

\section{Observations and Analysis} \label{observations}

The observations of SDSS~1352$+$4239 discussed in this paper are listed in
Table~\ref{obslog}.  

\subsection{Gemini GNIRS Observation}\label{gemini}

SDSS~1352$+$4239 was observed  using
GNIRS\footnote{http://www.gemini.edu/sciops/instruments/gnirs} on the
Gillett Gemini (North) Telescope using a standard cross-dispersed mode (the
SXD camera with the $31.7 \rm \, l/mm$ grating) and a $0{\farcs}45$
slit.  Eight 200-second exposures were made on 7 February 2015 in an
ABBA dither pattern.  Four 1-second exposures were made of the A0 star HIP
61471 at a similar airmass for telluric correction.  The data were
reduced using the IRAF {\it Gemini} package, coupled with the GNIRS XD
reduction scripts, in the standard manner for near-infrared spectra,
through the spectral extraction step.  For telluric correction, the
{\it Gemini} spectra of the source and the telluric standard star were
converted to a format that resembled IRTF SpeX data sufficiently that
the Spextool {\tt xtellcor} package \citep{cushing04, vacca03} could
be used.

\begin{deluxetable}{lcccc}
\tablewidth{0pt}
\tabletypesize{\small}
\tablecaption{Observations of SDSS~J1352$+$4239\label{obslog}}
\tablehead{
 \colhead{Observatory and Instrument} &
 \colhead{Date} &
 \colhead{Exposure (s)} &
 \colhead{Observed Frame Band Pass (\AA\/)} &
 \colhead{Resolution}\\  }
\startdata
SDSS & 2003 June 24 & 6300.0 & 3810--9189 & $100 \rm \, km\,
s^{-1}$ \\ 
{\it Gemini} (GNIRS) & 2015 February 7 & 1600.0 & 8263--25208 & $240 \rm \, km\,
s^{-1}$ \\
BOSS & 2016 April 5 & 8100.0 & 3628--10387 & $89 \rm \, km\,
s^{-1}$ \\
{\it APO} (Triplespec) & 2018 February 25 & 5280.0 & 9097--24704 & $80 \rm \, km\,
s^{-1}$  \\
\enddata
\end{deluxetable}

\subsection{APO Triplespec Observation}\label{apo}

SDSS~J1352$+$4239 was observed using 
Triplespec\footnote{https://www.apo.nmsu.edu/arc35m/Instruments/TRIPLESPEC/} \citep{wilson04}
on the Apache Point Observatory Astrophysical Research Consortium
3.5-meter telescope on 25 February 2018 under photometric conditions.
The 240-second observations were made in a standard  ABBA dither pattern and
split into two segments of 10 and 12 exposures.  Twenty 20-second exposures of the A0
star HIP 61471 were made before the first segment, and twelve
20-second exposures of the A0 star HIP 71172 were made after the
second segment. The  $1{\farcs}1$ slit was used.  The resolution was
measured using the night sky lines to be $80\rm\, km\, s^{-1}$ near
1.5 microns. 

The spectra were extracted in a standard manner using TripleSpecTool,
a modification of SpexTool \citep{cushing04, vacca03}. TripleSpecTool
uses the airglow emission lines for wavelength calibration. To account
for a very small amount of flexure, wavelength calibration solutions
were computed for each AB dither pair sequence of exposures.  The telluric correction was 
performed using the adjacent observation
of the A0 star \citep{vacca03}.  

The spectra were combined with the Gemini spectrum using a
flux-weighted average, where the variance was based on the deviations
of the spectrum around a best-fitting linear model to 21-pixel bins,
after first down-sampling the APO spectra to the Gemini resolution.
The combined spectrum is shown on the right panel in Fig.~\ref{spec_fig}.

\subsection{The SDSS and BOSS Observations and Merging the Spectra}
\begin{figure*}[!t]
\epsscale{1.0}
\begin{center}
\includegraphics[width=6.5truein]{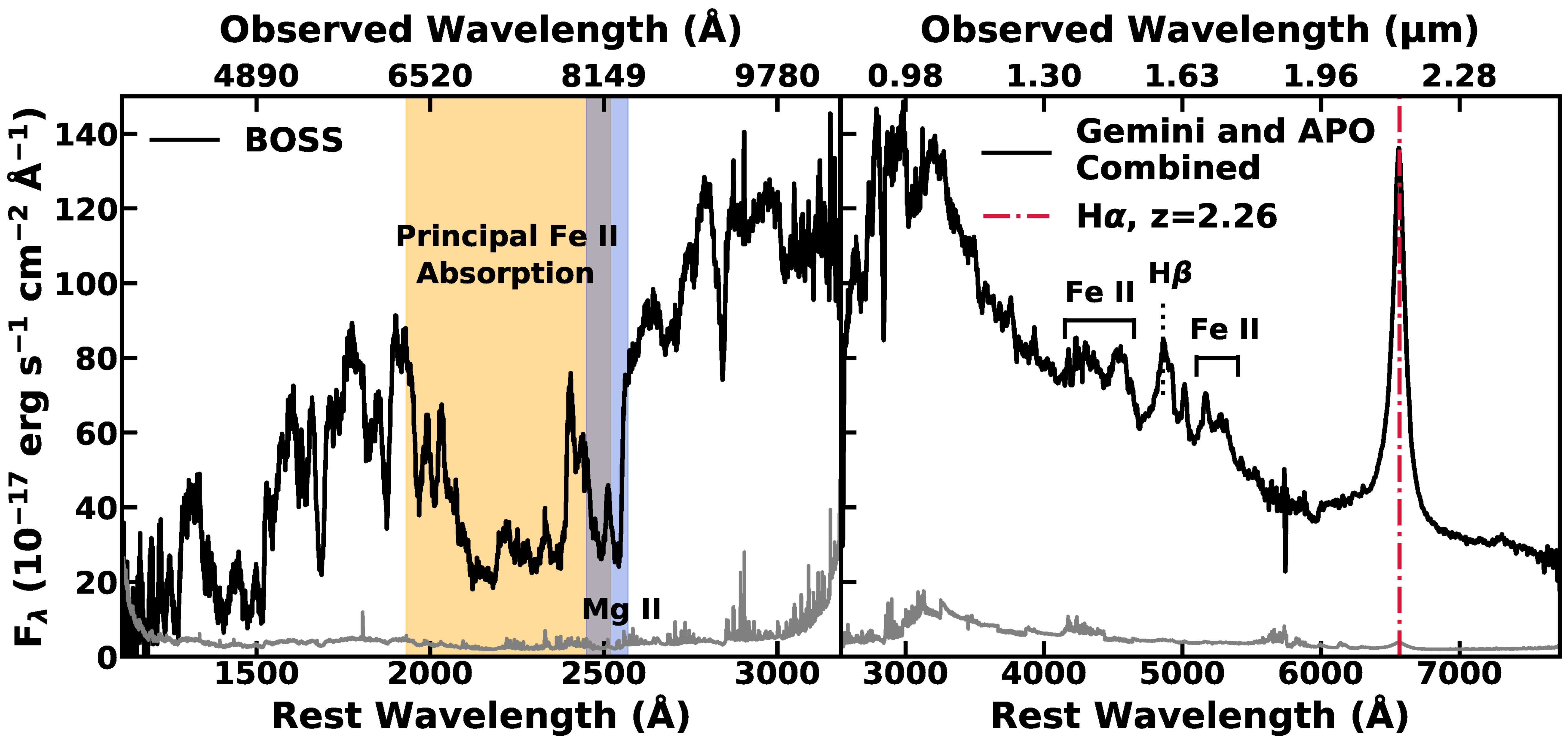}
\caption{The BOSS spectrum on the left shows an ``overlapping trough'' feature from the \ion{Fe}{2} absorption lines.
The main iron trough and \ion{Mg}{2} absorption features are marked on the left panel.
BOSS spectrum showed no strong emission features that could be used to estimate the redshift.
Therefore we used H$\alpha$ in the combined GNIRS+APO spectrum (right) to measure the redshift for SDSS~J1352+4239.
The flux level for the Gemini and APO combined spectrum has been corrected to match BOSS flux density.
The grey lines below the spectra show the uncertainties associated with the data. 
\label{spec_fig}}  
\end{center}
\end{figure*}

SDSS~J1352+4239 was observed by SDSS and by the Baryon Oscillation Spectroscopic Survey (BOSS) program.
We did not find any measurable flux offset or any strong evidence for spectral variability in the two spectra.
We chose to use the BOSS optical data from the SDSS archive because the data were taken closer to our near-infrared observations and the spectrum provides larger wavelength range coverage than the SDSS spectrum.
The BOSS and combined near-infrared Gemini and APO spectra are shown in Figure~\ref{spec_fig}.
We used the flux density of BOSS spectrum and the wavelength range between rest frame $\sim$ 3000 to $\sim$ 3100 \AA\/ to match and merge the optical BOSS and near-infrared Gemini and APO spectra.

\subsection{The Redshift}\label{correctz}

SDSS~1352$+$4239 was first cataloged in the SDSS Third Data Release
catalog \citep{schneider05}, where the redshift was listed as 2.0385.
Other published redshifts range from 2.000 \citep{meusinger00} to
2.049184 \citep{hewett10}.  The difficulty in estimating the redshift
occurs because there are no strong emission lines in the SDSS
spectrum.
A broad bump just longward of the \ion{Mg}{2} absorption was identified as \ion{Mg}{2} emission by \citet[][their Fig.\ 10]{trump06}.
On the other hand, the redshift of the absorption features is fairly obvious ($z=1.954$),
based on the characteristic pattern of \ion{Mg}{2} and \ion{Fe}{2}
absorption lines \citep[e.g.,][Fig.\ 12]{lucy14}.     

The redshift of SDSS~1352$+$4239 can be measured unambiguously from
the infrared spectrum.  We use H$\alpha$ because there are no
prominent [\ion{O}{3}] lines and H$\beta$ is blended with \ion{Fe}{2}
emission.
The line appears slightly asymmetric due to
  \ion{Fe}{2} emission so we fit it with two Lorentzian profiles.
The peak of the narrower one yields a redshift of $2.2639 \pm 0.0008$, 
$\sim 11$\% larger than any of  the previous estimated values,
implying that the outflow has a much larger velocity than previously
suspected.  

\subsection{The Black Hole Mass}\label{bhmass}

\begin{figure*}[!t]
\epsscale{1.0}
\begin{center}
\includegraphics[width=6.5truein]{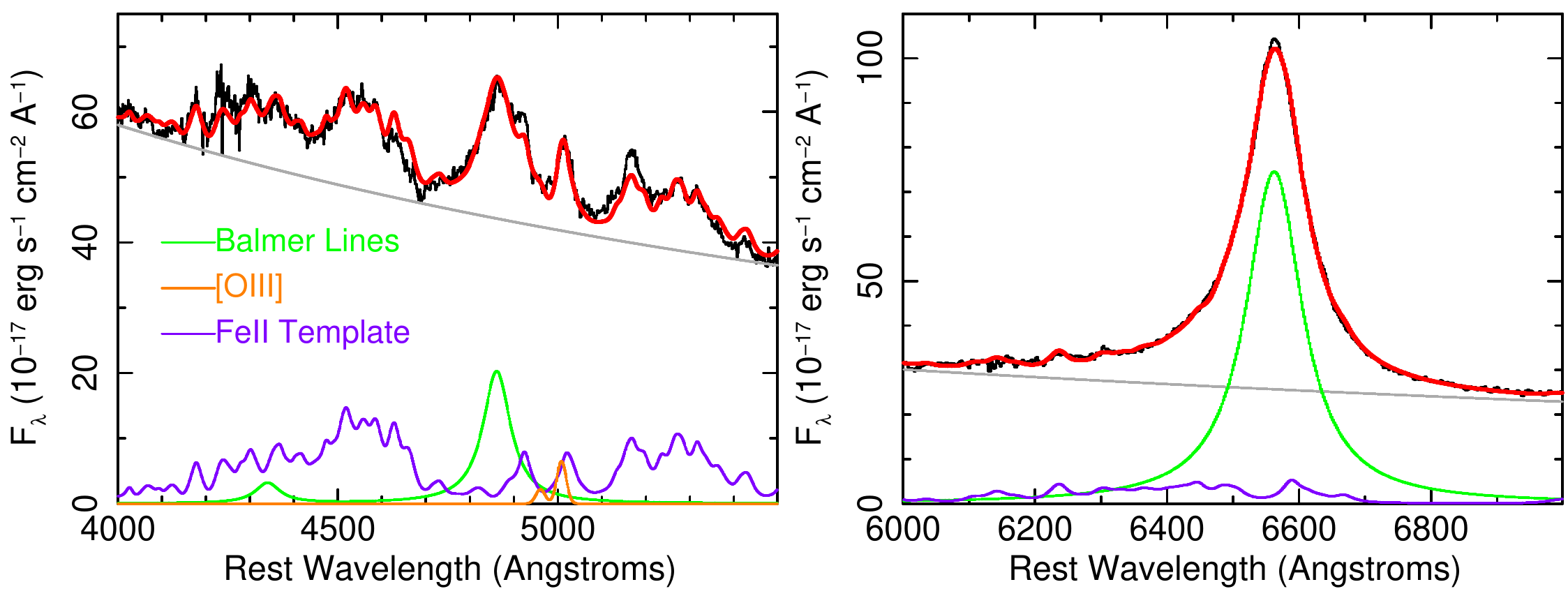}
\caption{The model fits to the combined Gemini and APO spectrum.  The
  left panel shows the bandpass that includes H$\beta$, and the right
  panel shows the bandpass that includes H$\alpha$. The strong
  \ion{Fe}{2} emission obscures the H$\beta$ line, so the two regions
  of the spectrum were fitted simultaneously, requiring that the FWHM
  of the Balmer lines to be equal.    \label{sdss1352_fits}}
\end{center}
\end{figure*}

We estimated the black hole mass using the H$\beta$ emission line.
Strong \ion{Fe}{2} emission is apparent throughout the rest-frame
  optical spectrum, and especially around H$\beta$.   We constrain the
  shape of H$\beta$ by simultaneously fitting Lorenzian profiles to
  each of H$\alpha$, H$\beta$, and H$\gamma$, and constraining their
  widths to be the same and their relative central wavelengths based on
  known wavelengths of these lines.
We used {\tt   Sherpa} for
spectral fitting \footnote{http://github.com/sherpa/sherpa/,
  http://cxc.harvard.edu/sherpa/} \citep{freeman01}.  The strong 
\ion{Fe}{2} emission was modeled using the catalog of \ion{Fe}{2}
emission lines obtained from I~Zw~1 \citep{vc04}.  No obvious
[\ion{O}{3}] lines are visible in the spectrum, but they are included
with a fixed width of $1500\rm \, km\, s^{-1}$ and variable position
and flux, with the 4960\AA\/ component constrained to have the same
width and fixed relative flux with respect to the 5008\AA\/
component.  The best-fitting model is shown in
Fig.~\ref{sdss1352_fits}.   

To determine the radius of the broad line region, we refer to
\citet{bentz13}, who find that $\log(R_{BLR})=K+\alpha \log[\lambda
  L_\lambda(5100)/10^{44}\rm \, erg\, s^{-1}]$. The continuum flux 
density at 5100\AA\/ was estimated from the combined Gemini and APO
spectrum to be $F_{5100} = 48.71 \times 10^{-17}\rm \, ergs\, s^{-1}\,
cm^{-2}$\AA\/$^{-1}$.  With the cosmological parameters used by  
\citet{bentz13} ($H_0=72\rm\,  km/s/Mpc$, $\Omega_M=0.27$,  and 
$\Omega_\Lambda=0.73$), we obtain a luminosity distance $D_L=18074\,\rm
Mpc$.  Using $K=1.527^{+0.031}_{-0.031}$ and
$\alpha=0.533^{+0.035}_{-0.033}$, we obtain an estimate of the radius
of the H$\beta$ emitting broad-line region of $1315^{+480}_{-340}$
light days corresponding to $1.1^{+0.4}_{-0.3}$ parsec. For reference, we
also calculated the 
location of the \ion{C}{4} emitting region using the equation given by
\citet[Equation (1)]{lira18}. We estimated the continuum flux density
at 1345\AA\/ to be $F_{1345} = 343.2 \times 10^{-17}\rm \, ergs\,
s^{-1}\,cm^{-2}$\AA\/$^{-1}$ after scaling the composite SED
\citep{richards06} to match the near-infrared (rest-optical)
photometry (\S~\ref{redlong}) and calculated the location of the
\ion{C}{4} emitting region of $199^{+436}_{-150}$ light days or
$0.17^{+0.37}_{-0.13}$ parsec. 

The model fit yields a FWHM of the Balmer lines of $4720\,  \rm
km\, s^{-1}$ for a Lorentzian profile.  We estimate the black hole
mass in the usual way.  We refer to \citet{collin06}, who provide
line-shape-based correction factors based on the ratio of the FWHM to
$\sigma_{line}$, where $\sigma_{line}$ is the line dispersion.  For a
Lorentzian profile, $FWHM/\sigma_{line} \Rightarrow 0$, and therefore
$f=1.5$.  We estimate that the black hole mass is $8.6\times 10^9 \rm
\, M_\odot$.

\section{Continuum Modeling and Spectral Energy Distribution}\label{contsedmod}

\begin{figure*}[!t]
\epsscale{1.0}
\begin{center}
\includegraphics[width=6truein]{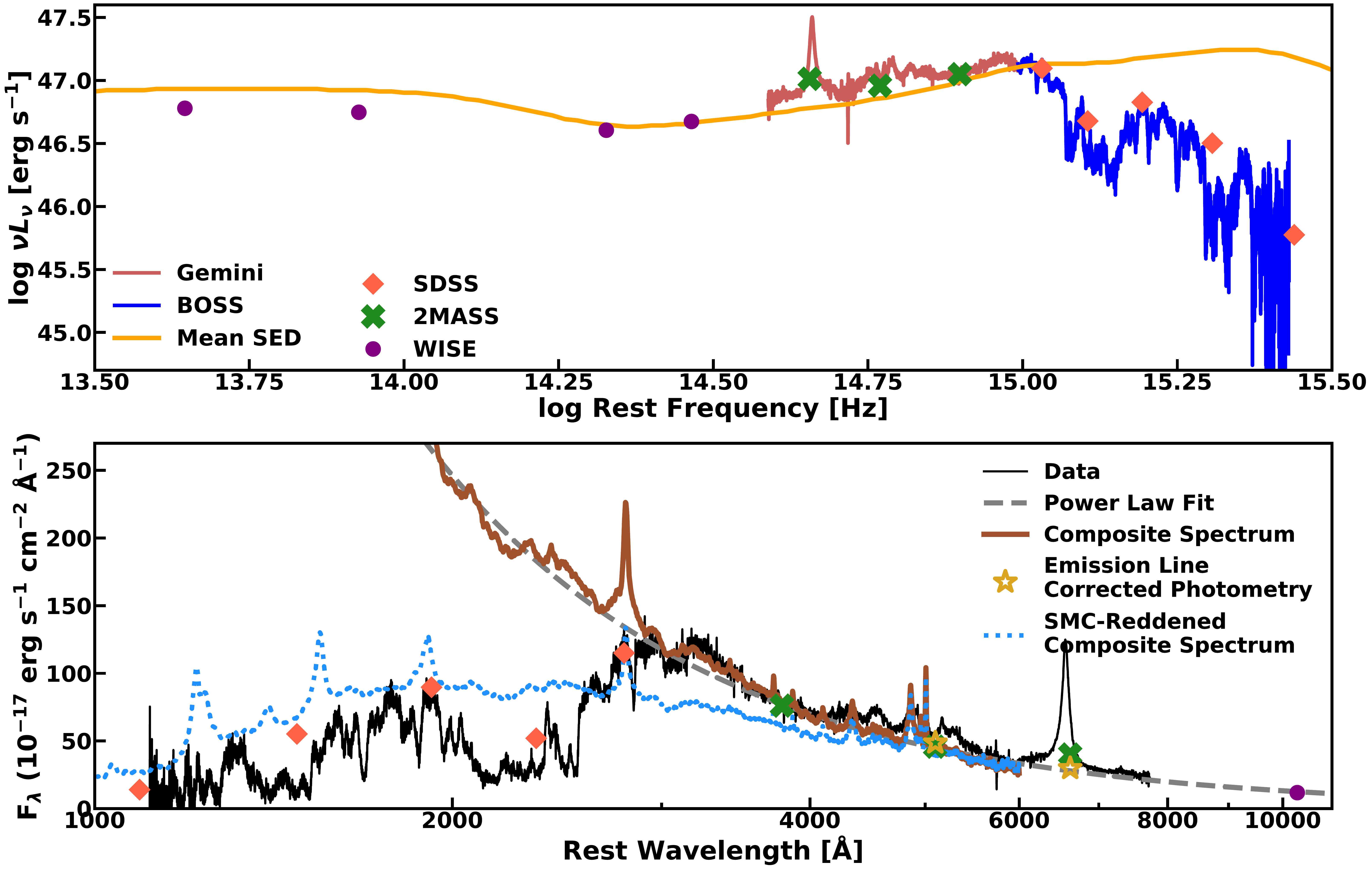}
\caption{SDSS~J1352+4239 is plotted with the mean SED from \citet{richards06} in the upper panel.
The lower panel shows the power law continuum fit to long wavelengths ($\lambda>\,3000\,\mathrm{\AA}$) and the composite spectrum from \citet{francis91}.
The H and K band photometry points have been corrected for the hydrogen line emission and iron emission using 2MASS filter functions \citep{cohen03} and iron emission templates created from the decomposition of the I~Zw~1 spectrum \citep{vc04}.
The SMC-reddened composite spectrum with $E(\bv)$=0.17, plotted in dotted blue in the lower panel, demonstrates that the SMC reddening curve fails to reproduce the continuum shape of SDSS~1352$+$4239.
While the observed and composite continuum shapes are similar longward of $\sim\,3000\,\mathrm{\AA}$, SDSS~1352$+$4239 diverges significantly at shorter wavelengths.
Because of the dramatic change in the SDSS~1352$+$4239 continuum shape at $\sim\,3000\,\mathrm{\AA}$ , we use a non-traditional reddening curve to model the continuum emission (\S~\ref{anored}).
\label{fig1}}  
\end{center}
\end{figure*}

\subsection{The Long-Wavelength Spectrum}\label{redlong}
SDSS~J1352+4239 shows a peculiar continuum shape compared to a typical quasar spectrum.
We used the composite quasar SED from \citet{richards06} and the composite spectrum from \citet{francis91} to analyze the shape of the underlying AGN continuum of the object using both the spectrum and the photometry from SDSS, 2MASS and WISE (Fig.~\ref{fig1}).
In Figure~\ref{fig1}, compared with the composite spectrum \citep{francis91}, the spectrum of SDSS~J1352+4239 is similar to a typical unreddened quasar at wavelengths longward of $\sim\,3000\,\mathrm{\AA}$.
In the infrared region, the shape of the SED of SDSS~J1352+4239 also resembles the mean quasar mid-infrared SED shape.
Because the continuum bluewards of the break shows a large difference in the slope, we analyzed the reddening and the slope of the continuum in the long wavelength region separately from the short-wavelength region.

\citet{krawczyk15} found that BAL quasars are redder than the non-BAL quasars, and that the SMC reddening curve (extinction curve derived from the Small Magellanic Cloud) fits BAL quasars well in most cases.
Therefore we used the SMC reddening law to measure the reddening in SDSS~J1352+4239.
We used the Markov Chain Monte Carlo code {\tt emcee}\footnote{http://dan.iel.fm/emcee/current/} \citep{emcee} to fit the SMC \citep{prevot84} reddened composite SED to the rest frame optical / near-infrared photometry points and found no evidence for reddening in the optical / near-infrared region of the spectrum ($E(\bv)\,<$ 0.002).

We also fit the optical / near-infrared part of the continuum using an SMC-reddened power law to get an estimate of the slope and reddening.
We measured a power law slope of $-1.82\,(\pm0.02)$, consistent with a mean spectral slope value for BALQs ($-1.83$, \citealt{krawczyk15}), and no reddening ($E(\bv)\,<$ 0.03) for the continuum from 1.4 \micron\/ to 3788 \AA.
Thus the object has a typical value of spectral slope and no evidence for reddening in the long wavelength region, despite significant reddening at shorter wavelengths.

To estimate the bolometric luminosity, we used the bolometric correction factor (BC) from \citet{galla07} who provide bolometric corrections for monochromatic luminosity at two different wavelengths.
The strong reddening in the spectrum is only seen at wavelengths shortward of $\sim$ 3000 \AA.
Therefore we used the monochromatic luminosity at 5100\AA\/ of SDSS~J1352+4239 (\S~\ref{bhmass}) and obtained the log bolometric luminosity of $48.0\pm0.2$ [erg $\rm s^{-1}$], with the uncertainties estimated from the uncertainties associated with the bolometric correction factor ($\rm BC=10.47\pm4.14$).

SDSS~J1352+4239 is among the most luminous quasars observed and it is considered a hyper-luminous quasar (i.e., quasars with $L_{\rm Bol} > 10^{47}$ erg $\rm s^{-1}$).
The bolometric luminosity of SDSS~J1352+4239 is comparable to the objects in the WISSH quasar sample \citep{bischetti17} where they focused on a sample of WISE/SDSS selected hyper-luminous quasars to study the power and the effect of the AGN feedback.
The mass accretion calculated from the bolometric luminosity, assuming the energy conversion efficiency ($\eta$) of 0.1, is 176 M$_{\odot}$ per year.
Compared with the black hole mass of $8.6\times 10^9 \rm
\, M_\odot$, SDSS~J1352$+$4239 is radiating at about 93\% of the Eddington limit.

\subsection{Anomalous Reddening}\label{anored}
As can be seen from Figure~\ref{fig1}, the shape of the continuum for SDSS~J1352+4239 is quite peculiar, but it is not unprecedented.
Among other BAL objects with anomalous reddening, Mrk 231 shows steep reddening in the near-UV to optical part of the continuum \citep[e.g.,][]{smith95,veilleux13}.
\citet{leighly14} fit the continuum in Mrk 231 and concluded that a Type Ia supernovae reddening curve \citep{goobar08} best describes the reddening behavior of Mrk 231.
\citet{jiang13} derived a reddening curve from IRAS~14026+4341 by comparing the object to a quasar composite spectrum and found that their reddening curve could be explained by a particular distribution of dust grain sizes (one lacking large grains, $a_{max}\ =\ 70\ \rm nm$).
However, in the case of WPVS 007 \citep{leighly09}, no particular grain distribution was able to model their anomalous reddening curve.

We tried using the reddening templates developed with WPVS 007 \citep{leighly09} and IRAS 14026+4341 \citep{jiang13} as well as the reddening model used for Mrk 231 \citep{leighly14} to model the break in the continuum shape.
However, none of the anomalous reddening models were able to appropriately model the continuum shape of SDSS~J1352+4239 because their slopes and the locations of sharp reddening increase did not match the continuum shape of SDSS~J1352+4239.\begin{figure*}[!t]
\epsscale{1.0}
\begin{center}
\includegraphics[width=3.5truein]{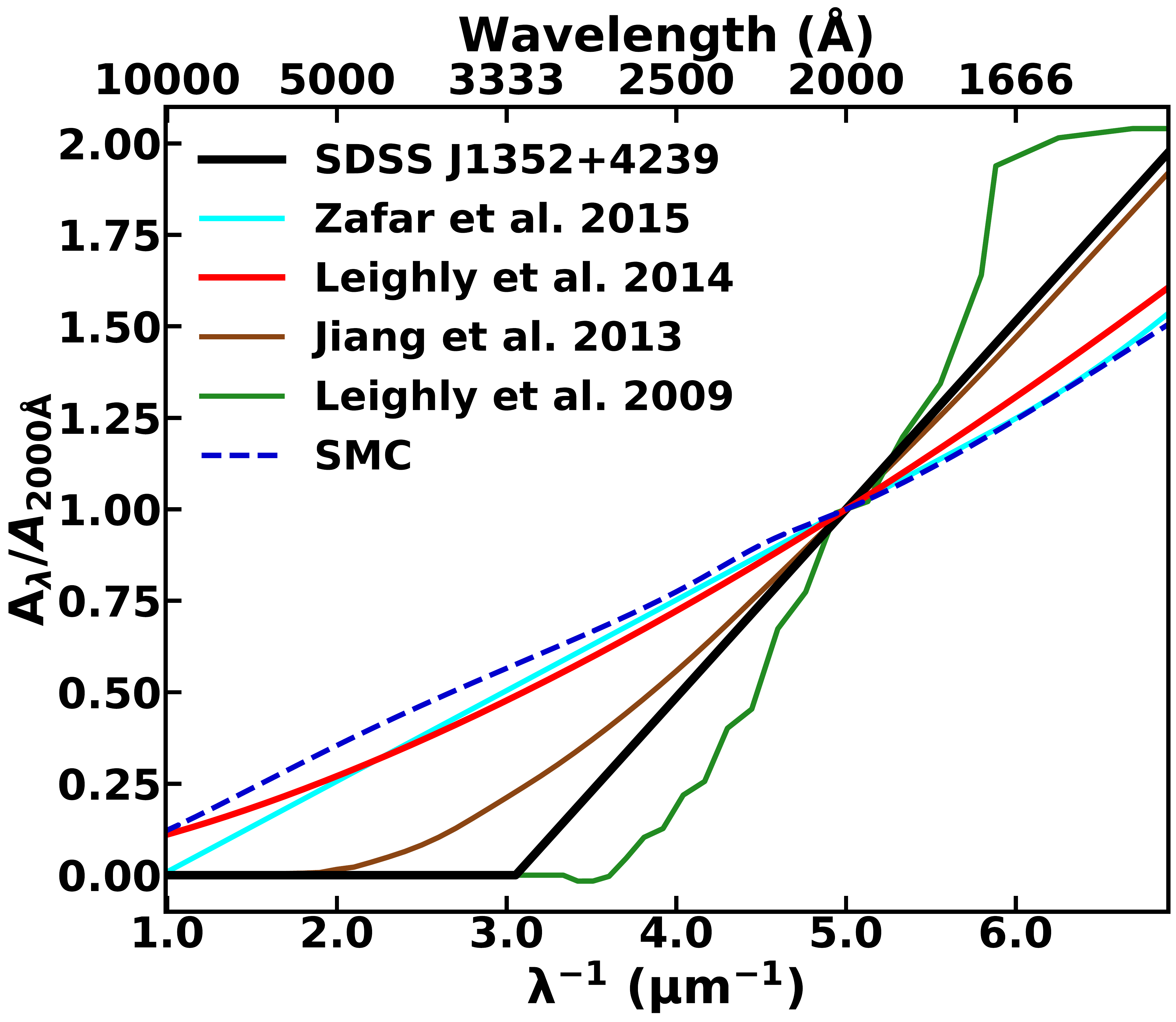}
\caption{The reddening curve for SDSS~J1352+4239 found from {\it SimBAL} fits using our model ($p=0.57\pm0.003,\ \lambda_{Break}=0.328\pm0.001\ (\micron)$) compared with other reddening curves developed for anomalous reddening.
The reddening curves have been normalized to $A_{\lambda}$ at 2000 \AA.
Anomalous reddening curves by \citet{leighly09} and \citet{jiang13} show different break wavelengths and slopes.
The SMC reddening curve and an empirical reddening curve derived from a sample of reddened quasars by \citet{zafar15} is also plotted ($A_V$ = 0.51) for comparison.
\label{red_fig}}  
\end{center}
\end{figure*}

Therefore, we developed a general anomalous reddening curve. Using the general reddening equation
A($\lambda) = 2.5 \log\{C (\lambda)/S (\lambda$)\} where S($\lambda$) is the reddened spectrum and C($\lambda$) is the intrinsic spectrum, our general reddening curve has the form of a power law.
\[A (\lambda\ (\micron)) = 
\begin{cases}
p(\frac{1}{\lambda}-\frac{1}{\lambda_{Break}}),\,(p>0) & \quad \text{if } \lambda\leq \lambda_{Break}\\
0 & \quad \text{if } \lambda > \lambda_{Break}
\end{cases}
\]
Our anomalous reddening curve generates reddening from a specified wavelength ($\lambda_{Break}$) to shorter wavelengths with A($\lambda)$ gradually increasing from zero, and therefore there is no reddening at long wavelength region as required.
The reddening equation requires two parameters: the slope of the curve ($p$) and a reddening starting wavelength ($\lambda_{Break}$).
Figure~\ref{red_fig} illustrates various reddening curves.
Our general reddening model provides excellent fits for other anomalously reddened BALQ spectra as well \citep{leighly_prep}.

To fit the shorter wavelength spectrum, we fixed the power law spectral slope to the value we found from the optical / near-infrared photometry fit, and only varied the two anomalous reddening parameters and the power law normalization to model the continuum with {\it SimBAL}.

\subsection{Modeling the Line Emission}\label{modelem}
Visual inspection of SDSS~J1352+4239 revealed that the object potentially has a weaker \ion{Mg}{2} emission and stronger iron emission compared with the typical AGN spectrum.
It is not possible to model the individual emission lines due to the heavy absorption features seen throughout the bandpass.
Instead, we constructed a set of broadband emission templates to model the emission lines.
It is well known that the ratio between the strengths of the prominent emission lines (e.g. \ion{Mg}{2}, \ion{C}{4}) and the strength of the iron emission differs from object to object \citep[e.g.,][]{sulentic00}.
Therefore, we created separate emission line templates for the iron emission and several other emission line templates for other emission lines so that our model can create the iron emission independently from other emission lines.
Mrk 493 is a narrow-line Seyfert with a strong \ion{Fe}{2} emission, making it a suitable target for AGN emission-line analysis.
It was observed by Hubble Space Telescope\footnote{PI: Park, ``A Definitive UV$-$Optical Template for Iron Emission in Active Galactic Nuclei'', program
  number 14744} to create a high resolution and good signal-to-noise ratio \ion{Fe}{2} template.
From this Mrk 493 spectrum, we derived empirical emission templates for the iron emission (the \ion{Fe}{2} pseudo-continuum) and for other emission lines (e.g. Ly $\alpha$, \ion{Si}{4}, \ion{C}{4}, \ion{C}{3}], \ion{Mg}{2}, Balmer lines) separately and used the extracted templates to model the emission features of SDSS~J1352+4239.

In order to separate the \ion{Fe}{2} emission from the other emission lines in the Mrk 493 spectrum, we used {\tt Sherpa} to model the spectrum using a power law, existing \ion{Fe}{2} templates (\citet{vc04}: 4000 \AA\/ $\lesssim\,\lambda_{rest}\,\lesssim$ 7000 \AA\/, \citet{lm06}: 2000 \AA\/ $\lesssim\,\lambda_{rest}\,\lesssim$ 3000 \AA\/ and \citet{leighly11}: 3000 \AA\/ $\lesssim\,\lambda_{rest}\,\lesssim$ 4000 \AA\/) and Gaussian line profiles for all other emission lines present in the spectrum.
We then subtracted the Mrk 493 spectrum by the emission-line models consisting of only the non-\ion{Fe}{2} emission lines and power law continuum to obtain the \ion{Fe}{2} emission templates.
Separate emission templates for other major emission lines were made from the non-\ion{Fe}{2} emission line component of the same model.
We merged the resulting \ion{Fe}{2} emission templates together to create a single broadband emission template (1500 \AA\/ $\lesssim\,\lambda\,\lesssim$ 7500 \AA\/).
We did not attempt to do the same for the non-\ion{Fe}{2} emission line templates to allow {\it SimBAL} more flexibility in fitting the major emission-line features so that each templates can be scaled to their own independent normalization coefficients.
The final emission-line templates consist of a single full wavelength range template for \ion{Fe}{2} emission lines and 4 emission templates divided in wavelength sections mentioned above for the non-\ion{Fe}{2} AGN emission lines.

\section{Best-Fitting Model}\label{bestmodel}
We created a complex spectral model for SDSS~J1352+4239 to extract the physical properties of the outflow.
Our best-fitting model is made of 4 major components including two absorbing components.
The continuum and line emission were modeled by a power law and emission line templates described in \S~\ref{modelem}.
A scattered non-absorbed continuum emission component was added to the model to produce the peculiar non-black saturation shape under the iron trough.
Reddening was applied to all components using an anomalous reddening model discussed in \S~\ref{anored}.
We first discuss the main blueshifted absorption-line component in \S~\ref{modelcomp}, then explore the necessity of the scattered light component in \S~\ref{scatter} and a zero-velocity absorption component in \S~\ref{redcomp}.
The results are summarized in Table~\ref{param_table}

\begin{figure*}[!t]
\epsscale{1.0}
\begin{center}
\includegraphics[width=7truein]{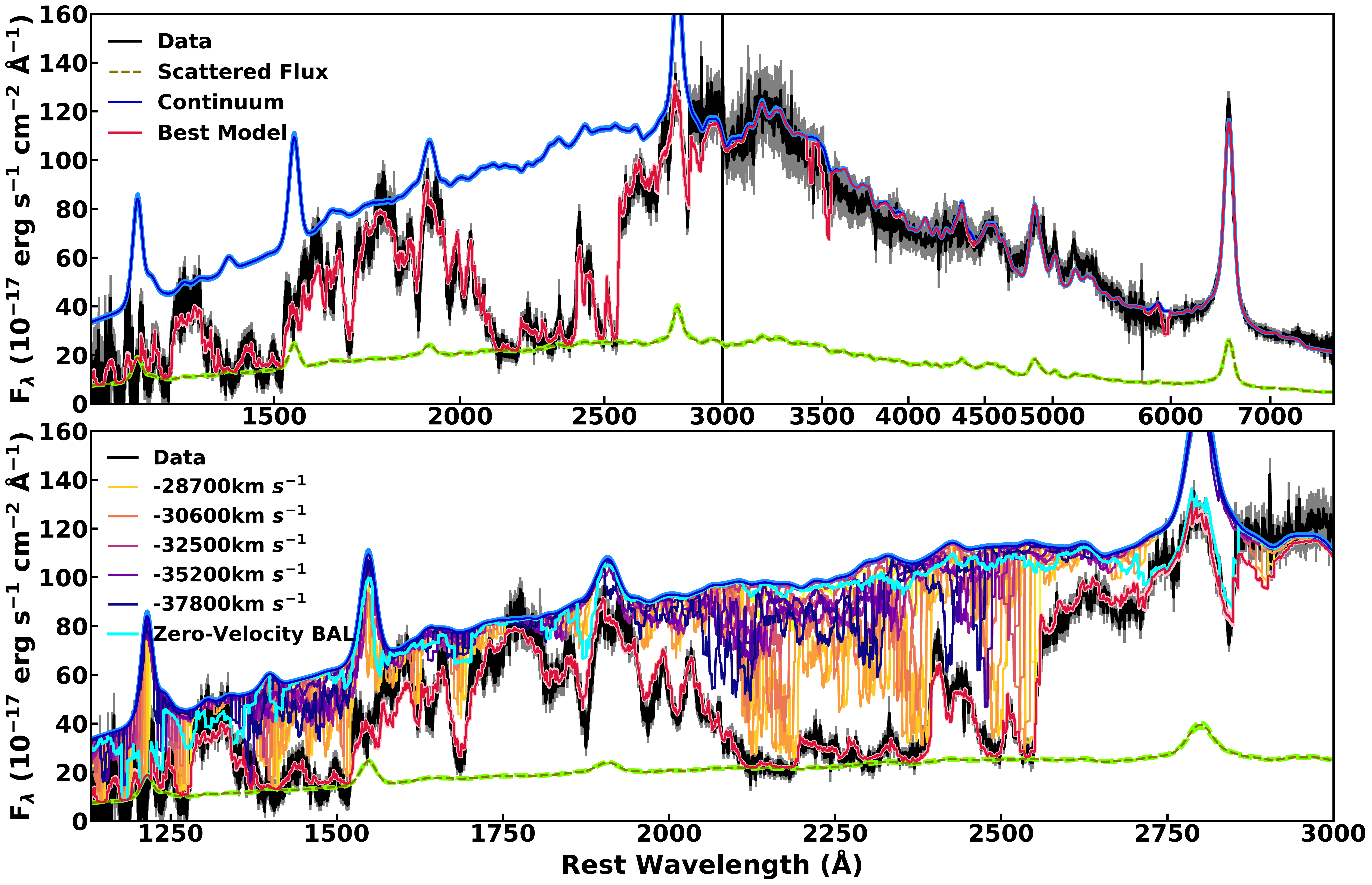}
\caption{Upper panel: Our best fitting model described in \S~\ref{bestmodel}.
Lower panel: Decomposition of ten tophat bins is shown in different colors (from yellow to navy); the zero-velocity BAL component is plotted in cyan.
The velocities of five of the ten tophat bins for the main complex are labeled on the figure.
Each bin in the absorption complex creates an absorption feature at a different velocity.
The combination of 10 bins create the full trough and we harvest the information about the physical parameters of the outflow as a function of velocity.
\label{fullbestfig}}  
\end{center}
\end{figure*}

The model is given by:
\begin{align*}
f_{model}=Reddening\times\{(f_{Continuum}+f_{Line Emission})\times I_{High-Velocity}\times I_{Zero-Velocity}+f_{Scattered\,Flux}\}
\end{align*}
where $f(\lambda)$ is the flux from each component and the final model and $I(\lambda)$ is the normalized flux ($I/I_0$) from each absorption component. 
Figure~\ref{fullbestfig} shows the best fit model of SDSS~J1352+4239.

Depending on the geometry and the angular size scale of the BAL outflowing cloud, the covering fraction for the accretion disk and the line-emitting gas (broad line region, BLR) can be different.
\citet{leighly18b} demonstrated how {\it SimBAL} can be used to test the scenarios where the outflowing cloud has multiple covering fractions for different AGN components.
We tested both two-covering models where the covering-fraction parameters for the line emission and the continuum emission were allowed to differ and single-covering models and concluded that there is no strong evidence for a different covering fraction for emission lines and continuum emission in SDSS~J1352$+$4239.
Therefore we used a model with a single covering fraction for both emission components.

The tophat accordion model provided an exceptional fit of the complex velocity structures of the trough in SDSS~J1352$+$4239, and yielded the physical parameters of the outflows as a function of velocity (Fig.~\ref{fit_par}).
We fit the high-velocity troughs with a 10-bin tophat model with an additional 7-bin tophat model for the zero-velocity absorption feature we identify near the \ion{Mg}{2} emission lines (\S~\ref{redcomp}).
\citet{leighly18a} explored the dependence on number of bins and concluded that the number of bins does not change the result of the fit except when too few bins were used, and that there were no significant differences between the results obtained with models with different number of bins.
We experimented with 7, 10 and 15-bin tophat accordion models and found that 10 bins were sufficient to model the complex.
Ten bins span a velocity range from $\sim-38000 \rm \, km\, s^{-1}$ to $\sim-28000 \rm \, km\, s^{-1}$ with the total velocity width of $\sim 10000 \rm \, km\, s^{-1}$ (Fig.~\ref{bestmodel}).

The physical parameters and the derived outflow properties for the high velocity trough and zero-velocity component (\S~\ref{redcomp}) as well as for each group are reported in Table~\ref{param_table}.
The main blueshifted trough in SDSS~J1352+4239 was modeled with a 10-bin tophat accordion model where the bins were divided into two groups with a single ionization parameter and density for all bins in each group as described in \S~\ref{modelcomp}.
The values for $\log U$, $\log n\rm\,[cm^{-3}]$, $\log N_H\,-\,\log U\rm\,[cm^{-2}]$ and $\log a$ were directly taken from the the physical fit parameters of the best-fitting model.
The hydrogen column density values that have been corrected for the partial coverage with $\log a$ and the outflow properties (e.g., $\log\dot M$, $\log\ L_{KE}$) have been calculated from the aforementioned fit parameters.
For $\log N_H\,-\,\log U\rm\,[cm^{-2}]$, $\log a$ and $\log N_H\rm\,[cm^{-2}]$, the ranges reflect the values we found for the individual bins.
Total $\log N_H$ for the groups are also reported.
Uncertainties for each parameter were calculated from the posterior probability distributions of the MCMC chain.
We did not attempt to model the posterior distribution (e.g., Gaussian distribution), instead we calculated the median, 1$\sigma$, 2$\sigma$ and 3$\sigma$ values directly from the posteriors.
The uncertainties reported in the Table~\ref{param_table} represents 95\% confidence regions.
A global covering fraction ($\Omega$) of 0.2 was used for the calculations and further discussion of this parameter can be found in \S~\ref{postpro}.

\begin{deluxetable}{lcccc}
\tablewidth{0pt}
\tabletypesize{\scriptsize}
\tablecaption{Physical Parameters and Derived Outflow Properties from the Best-Fitting {\it SimBAL} Model\label{param_table}}
\tablehead{\\
 \colhead{Outflow Properties} & 
 \colhead{Higher Velocity Group} & 
 \colhead{Lower Velocity Group} &
 \colhead{High-Velocity Total\tablenotemark{a}} &
 \colhead{Zero-Velocity Component}}
\startdata
\multicolumn{5}{c}{Physical Parameters} \\
\hline
$v_{outflow}\rm\ (km\ s^{-1})$\tablenotemark{b}&$-38000$ to $-33000$&$-33000$ to $-28000$&$-38000$ to $-28000$&$-8900$ to $6700$\\
$\log U$&$0.82^{+0.07}_{-0.12}$&$-0.56^{+0.07}_{-0.08}$&-&$-2.8$ to $1.8$\tablenotemark{b}\\
$\log n\rm\,[cm^{-3}]$&$6.12^{+0.12}_{-0.07}$&7.43$^{+0.09}_{-0.07}$&-&$<5.0$\tablenotemark{e}\\
$\log N_H\,-\,\log U\rm\,[cm^{-2}]$\tablenotemark{b}&23.0--23.16&23.13--23.17&-&21.9--23.0\\
$\log a\tablenotemark{b}\tablenotemark{c}$&0.91--1.9&0.38--1.13&-&$-0.58$ to 1.92\\
\hline
\multicolumn{5}{c}{Derived Outflow Properties} \\
\hline
$\log N_H\rm\,[cm^{-2}]$, per bin\tablenotemark{b}\tablenotemark{c}&22.03--22.85&21.41--22.06&-&18.31--21.82\\
$\log N_H\rm\,[cm^{-2}]$, total\tablenotemark{d}&$23.11^{+0.07}_{-0.06}$&$22.57^{+0.06}_{-0.07}$&$23.22\pm0.05$&$21.85^{+0.05}_{-0.06}$\\
$\log R$ [pc]&0.97$^{+0.05}_{-0.04}$&1.0$\pm0.02$&0.93--1.02&$>1.0$\tablenotemark{e}\\
$\log\dot M\ \rm[M_\odot\,yr^{-1}]$\tablenotemark{f}&$3.41^{+0.04}_{-0.05}$&$2.81^{+0.06}_{-0.07}$&$3.51\pm0.04$&-\\
$\log \dot P$ [dyne]\tablenotemark{f}&$38.77^{+0.04}_{-0.05}$&$38.08^{+0.06}_{-0.07}$&$38.85\pm0.04$&-\\
$\log\ L_{KE}\rm\ [erg\ s^{-1}]$\tablenotemark{f}&$48.04^{+0.04}_{-0.05}$&$47.25^{+0.06}_{-0.07}$&$48.1\pm0.04$&-\\
\enddata
\tablenotetext{a}{The values are the combined result of the left two columns.}
\tablenotetext{b}{The range of values estimated from the multiple bins is reported.}
\tablenotetext{c}{Large value of $\log a$ corresponds to small covering fraction}
\tablenotetext{d}{Covering fraction weighted values are reported (\S~\ref{modelcomp}).}
\tablenotetext{e}{Zero-velocity component is located at a larger distance than the main high velocity component (\S~\ref{geometry}).}
\tablenotetext{f}{The global covering fraction $\Omega=0.2$ was used \citep[e.g.,][]{hewett03}, and further discussion of $\Omega$ can be found in \S~\ref{postpro}.}
\end{deluxetable}

\subsection{The High-Velocity Component}\label{modelcomp}
The 10 bins for the main high-velocity trough were grouped into two sets with each group having a single density and ionization parameter.
Our initial investigation with {\it SimBAL} models revealed that the bins at higher velocities and at lower velocities have clear differences in their physical parameters, primarily in thier densities.
Subsequently, we found that the two density groups also had different characteristic ionization parameters.
Therefore, we assigned a single ionization parameter and density to each group.
\begin{figure*}[!t]
\epsscale{1.0}
\begin{center}
\includegraphics[width=3.5truein]{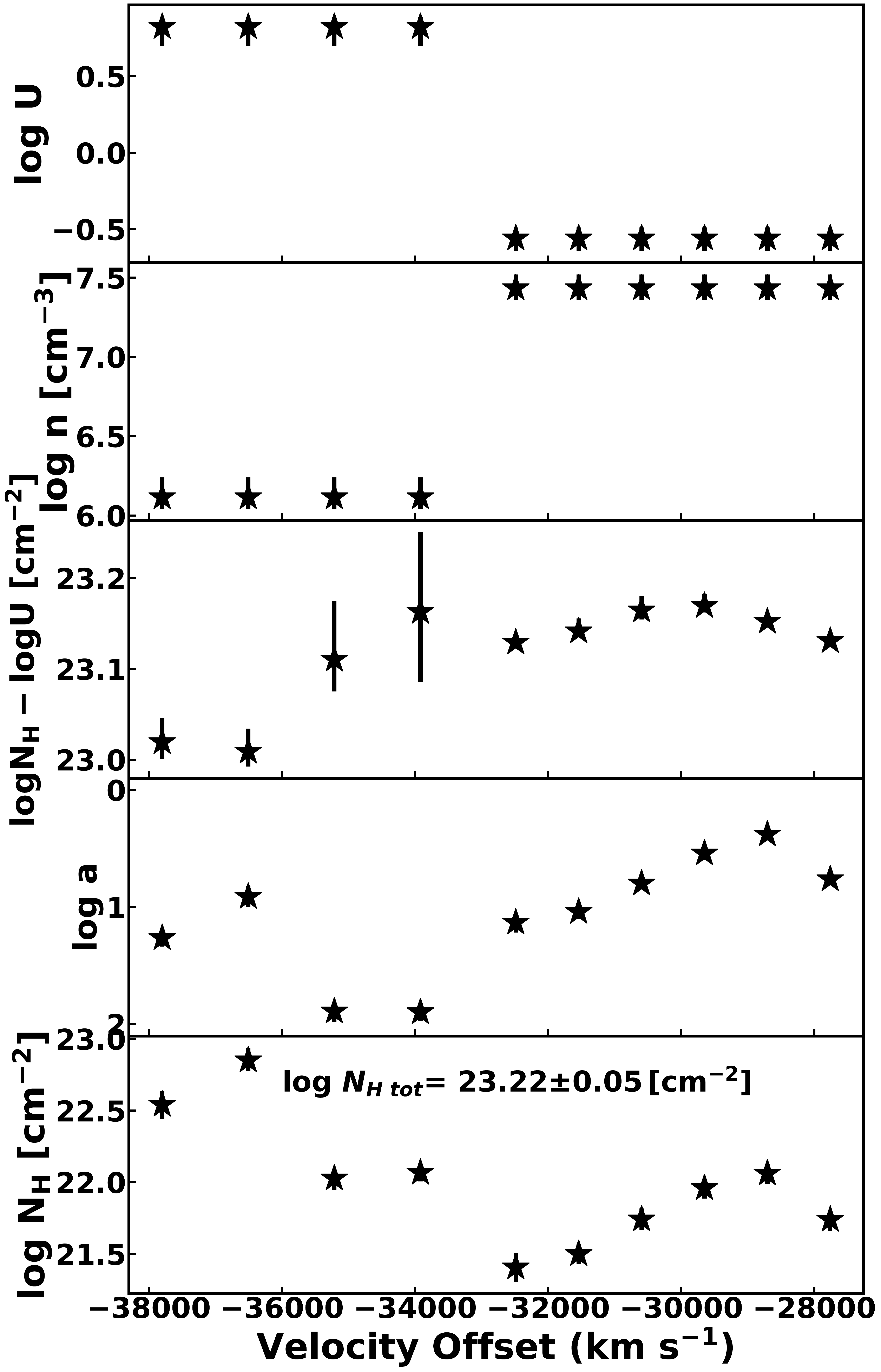}
\caption{Physical parameters as a function of velocity with error bars representing 95\% confidence regions.
The parameters plotted in the top 4 panels were directly fitted with {\it SimBAL} and in the bottom panel, the hydrogen column density values (log $N_H$), corrected for the covering fraction from each bin, were calculated from log $U$, $\log N_H - \log U$ and log $a$.
The total log $N_H$ value for the outflow, calculated from adding the hydrogen column density values from all 10 bins is also reported in the bottom panel.
The two groups ($-38000\sim-33000\,\rm \, km\, s^{-1}$ and $-33000\sim-28000\,\rm \, km\, s^{-1}$) are constrained to each have the same density and ionization parameter (top two panels), while the log $N_H$ $\sbond$ log $U$ parameter and the covering fraction parameter (lower log $a$ values indicate higher covering fraction) were allowed to vary independently for each bin.
The highest covering fraction (lowest log $a$ value) occurs around $\sim-30000\,\rm \, km\, s^{-1}$ and the column density parameter log $N_H$ $\sbond$ log $U$ also peaks around the same velocity.
This shows that most of the opacity is generated near this velocity (see also Fig.~\ref{detail_fig}).
\label{fit_par}}  
\end{center}
\end{figure*}

\ion{Fe}{2} has a plethora of excited state levels, ranging from low level excited states (0-0.12 eV) as well as high levels ($>$2.89 eV), making the strengths of the excited state \ion{Fe}{2} lines very density sensitive \citep[e.g.,][]{lucy14}.
\ion{Fe}{2} ions are populated deep in the photoionized cloud away from the incoming radiation because the ionization potentials to create \ion{Fe}{2} ions is relatively low (7.9 eV). 
Therefore \ion{Fe}{2} ions require a large column density to be significant (column density reaching beyond hydrogen ionization front), otherwise most of the iron atoms will be in a higher ionization state than \ion{Fe}{2}.
Thus the presence of the excited state \ion{Fe}{2} lines along with other low ionization lines (e.g., \ion{Mg}{2}) helps {\it SimBAL} to constrain both the density and the thickness of the outflowing gas.
We see in Figure~\ref{fullbestfig} not only how all 10 bins model the trough together in combination but also how each tophat bin creates a large number of absorption lines.
Together the physical parameters at each velocity can be constrained.

\begin{figure*}[!t]
\epsscale{1.0}
\begin{center}
\includegraphics[width=7truein]{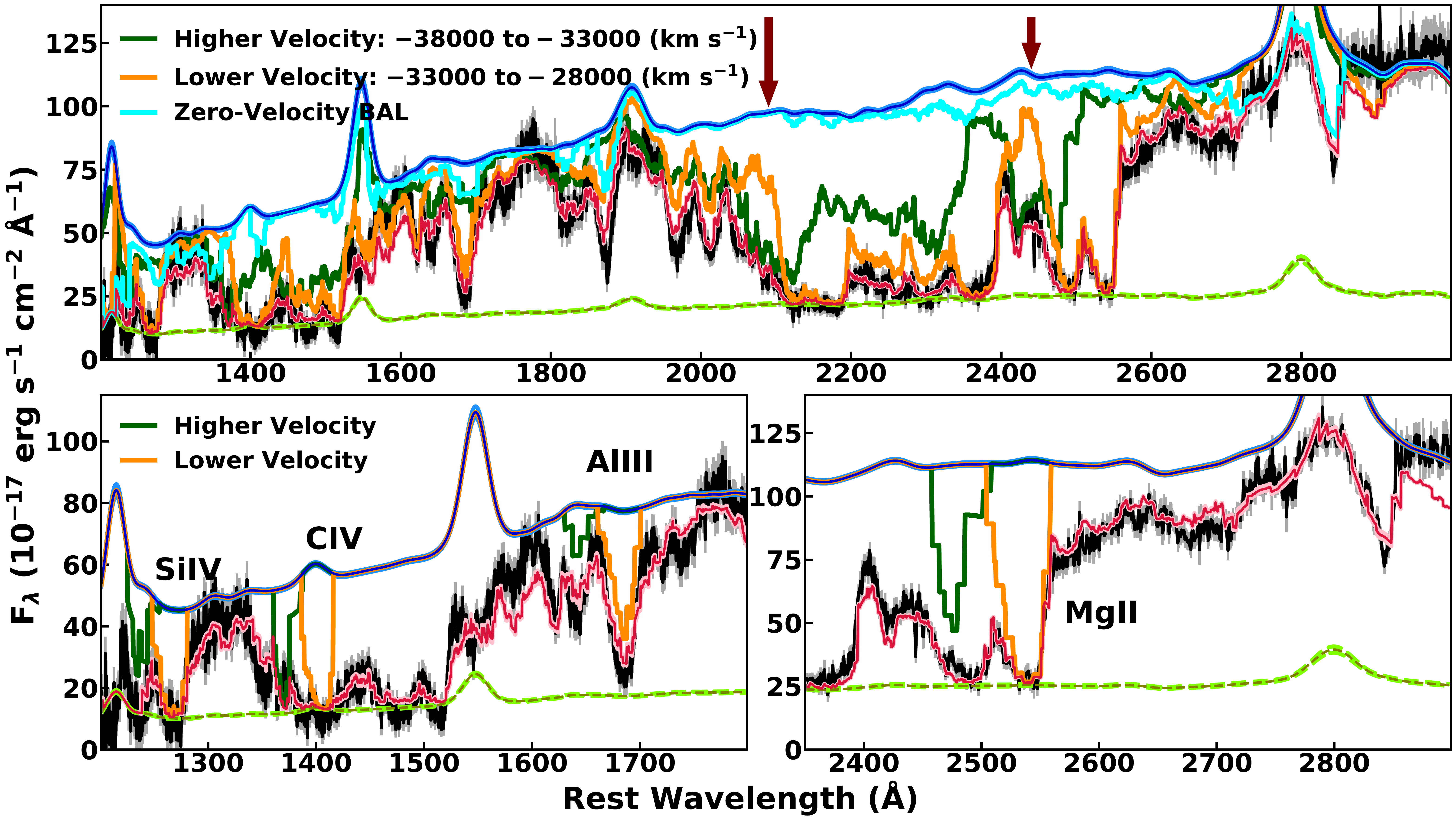}
\caption{The top panel shows the two models generated from combining only the higher and lower velocity bins in dark green and orange, respectively.
The regions where the higher velocity group plays a significant role in producing sufficient opacity to model the trough are marked with arrows in the top panel.
The bottom two panels show how some of the common BAL absorption lines (\ion{Si}{4}, \ion{C}{4}, \ion{Al}{3}, \ion{Mg}{2}) have been modeled by the higher velocity group and the lower velocity group.
The best-fitting model, continuum and the scattered flux component are plotted in same colors as Fig.~\ref{fullbestfig}.
\label{detail_fig}}  
\end{center}
\end{figure*}

Figure~\ref{fit_par} shows the outflow physical parameters as a function of velocity.
We found the high velocity part of the outflow has lower density ($\log n\sim6.12\ \rm[cm^{-3}]$) and higher ionization ($\log U\sim0.82$) than the lower velocity group ($\log n\sim7.43\ \rm[cm^{-3}]$, $\log U\sim-0.56$).
The large combination parameter (log $N_H$ $\sbond$ log $U$) of $\sim$ 23.1 $\rm[cm^{-2}]$ reflects the significant opacity from \ion{Fe}{2} ions that we see in the data.
The covering fraction parameter ($\log a$) changes strongly with the velocity and the bottom panel in Figure~\ref{detail_fig} shows how the shape of the opacity profile of the absorber closely follows the shape of $\log a$.
Moreover, the large covering fraction (low log $a$) and high log $N_H$ $\sbond$ log $U$ parameter found near $\sim-29000 \,\rm \, km\, s^{-1}$ indicates that a large amount of opacity is concentrated around that velocity region in the outflow.
Similarly, \citet{leighly18a} also found a ``concentration'' region in their {\it SimBAL} model of SDSS~J0850+4451, i.e., an enhancement in column density for a few of the bins in their 11-bin tophat model.
By summing the hydrogen column density values weighted by the covering fraction from all 10 bins, each calculated from the $\log U$ parameter, log $N_H$ $\sbond$ log $U$ parameter, and covering-fraction parameter (log $a$) per bin \citep[$\log N_H=(\log N_H\sbond\log U)+\log U-\log(1+10^{\log a})$][]{arav05,leighly18a,leighly18b}, we estimated a covering fraction weighted total hydrogen column density of log $N_H=23.22\pm0.05\,[\rm cm^{-2}]$ (95\% confidence errors, bottom panel in Figure~\ref{fit_par}).

Figure~\ref{detail_fig} shows how the two tophat groups model the wide absorption feature.
The higher velocity component contributes less opacity than the lower velocity component; however, the lower velocity component alone cannot produce the wide trough we see in the data.
The lower velocity component has gaps between $\sim 2450\,\rm\AA$ and $\sim 2600\,\rm\AA$, and near $\sim 2100\,\rm\AA$ where the \ion{Fe}{2} and other iron peak ions in the high-excited states are expected to be the main source of the opacity.
The problem is that the lower velocity component cannot produce enough opacity in those regions without creating a deep absorption feature near $\sim 2600\,\rm\AA$ that is not present in the spectrum.
Therefore the higher velocity group (with distinct values for the density and ionization parameter) was needed to fill in the gaps in the trough where the lower velocity component did not produce enough opacity to complete the absorption feature (arrows in Fig.~\ref{detail_fig}).

In Figure~\ref{detail_fig}, we also see that the concentration of opacity and strong absorption contribution from the lower velocity component, as expected from Figure~\ref{fit_par}, and the shape of the absorption profile for an individual transition (dark green and orange lines in the lower panel) closely follows the shape of the covering fraction parameter.
The blended lines in the main trough are nearly saturated even with the partial covering; the flux at the bottom of the trough is mainly modeled by the scattered light component.

\subsection{The Scattered Light Component}\label{scatter}
SDSS~J1352+4239 shows an extreme case of non-black saturation in the main trough where the emission at the bottom of the trough increases as a function of wavelength and contains a significant amount of flux.
Non-black saturation of BAL features is very common and is thought to originate from the BAL outflow not entirely covering the continuum sources, which includes the accretion disk continuum and broad emission line features \citep[e.g.,][]{barlow97}.
Continuum scattering is not uncommon in BALQs, and it is known from spectropolarimetry that frequently the troughs are highly polarized indicating an origin in scattered light \citep[e.g.,][]{cohen95,ogle99}.
The shape of the offset found under the trough in SDSS~J1352+4239 suggests that this component is scattered light from the accretion disk continuum and line emission with the wavelength dependence created by the reddening.
We modeled the scattered light component by multiplying the scattering fraction parameter by the emission model consisted of the sum of the reddened power law continuum and line emission and added this component to the absorbed emission model: $$f_{Scattered\,Flux}(\lambda)=(f_{Continuum}(\lambda)+f_{Line Emission}(\lambda))\times Scattering\, Fraction.$$
The reddening of the scattered flux is assumed to be the same as the continuum reddening, and we assume that the scattered light is not absorbed by the wind.
Our best model creates the underlying emission feature with a scattering fraction of $\sim$ 29$\pm$0.5\%.
This value is large but comparable to the scattering fraction of $> 20\%$ found in IRAS~13349$+$2438 by \citet{lee13}.
A large scattering fraction suggests that SDSS~J1352$+$4239 may be highly polarized.
Considering the amount of polarization depends both on the geometry of the scattering source and the scattered fraction, SDSS~J1352$+$4239 may exhibit polarization less than this value.
Previous spectropolarimetry observations of BALQSOs revealed polarization reaching greater than $\sim10\%$ in some objects \citep[e.g.,][]{brotherton97,ogle99}.

To test the necessity of the scattered flux component, we fit the data with a model that does not include it.
The model fails to match the shape around $\sim 2100 - 2200 \,\rm\AA$, creating a deeper \ion{Fe}{2} trough.
\begin{figure*}[!t]
\epsscale{1.0}
\begin{center}
\includegraphics[width=6truein]{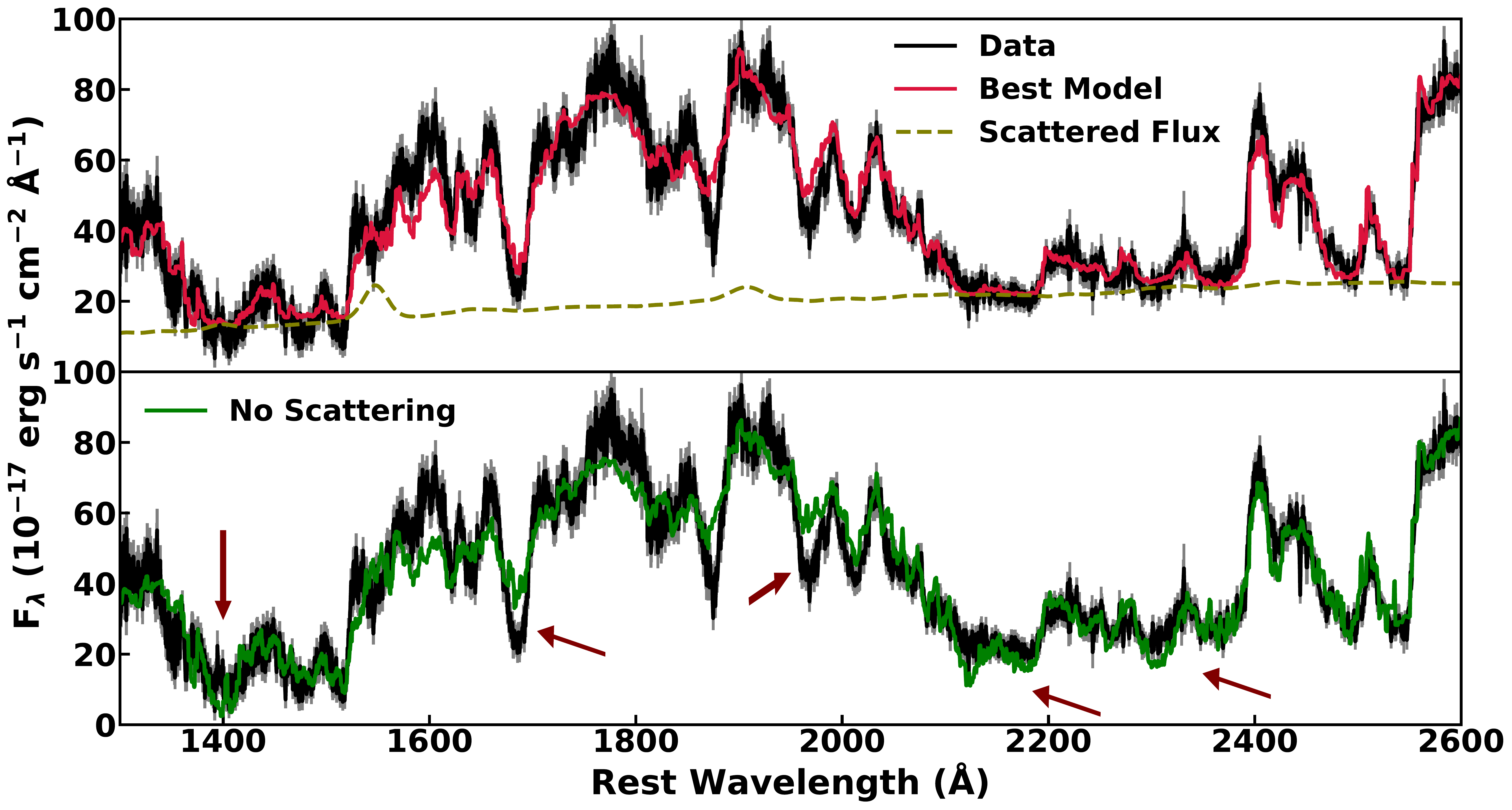}
\caption{The top panel shows the data and the best fit model that has the scattered flux component in it. The bottom panel shows a model that does not have the scattered light component.
The scattered light component is clearly necessary to create an appropriate trough shape.
\label{fig2}}  
\end{center}
\end{figure*}
Figure~\ref{fig2} shows the comparison between the best fitting model and the model without the scattered component.
Further discussion of possible origins of the scattered light is given in \S~\ref{geometry}.

\subsection{The Zero-Velocity Component}\label{redcomp}
We found a single prominent absorption feature between 2800\AA\/ and 2850\AA\/ that was not modeled with the blueshifted components (Fig.~\ref{fullbestfig} and \ref{detail_fig}).
We identified this feature as \ion{Mg}{2}$\lambda\lambda 2796,2803$ lines with near zero velocity offset and modeled it with a separate group of tophats bins.
Seven tophat bins for the zero-velocity component span a velocity range from $\sim-8900 \rm \, km\, s^{-1}$ to $\sim6700 \rm \, km\, s^{-1}$ with the total velocity width of $\sim 15000 \rm \, km\, s^{-1}$.
The zero-velocity component seems to be most prominent in the \ion{Mg}{2} lines and this doublet is the only feature that is not blended significantly with the high-velocity lines.
Our model also found the low-ionization lines  \ion{Al}{3}$\lambda\lambda 1854,1862$ and \ion{Al}{2}$\lambda 1670$ from the zero-velocity component to be present as shallow features in the spectrum at $\sim 1880\,\rm\AA$ and $\sim 1670\,\rm\AA$ with the \ion{Al}{2} line being the shallower of the two.

Notably, we find no strong evidence for high-ionization absorption lines such as \ion{Si}{4}$\lambda\lambda 1402,1393$  and \ion{C}{4}$\lambda\lambda 1548,1550$ from the zero-velocity component in the data.
That is, the high-velocity component alone produces enough opacity to match the data in the regions where the high-ionization lines from the zero-velocity component are expected to appear.
This is very unusual since \ion{Al}{3} and \ion{Mg}{2} are always accompanied by high-ionization lines \citep[e.g.,][]{voit93}.
Moreover, the high ionization conditions that produce larger \ion{Al}{3} opacity than \ion{Al}{2} opacity for the zero-velocity component also predicts significant high-ionization lines.

We suspect that the gas cloud for the zero-velocity component is illuminated by continuum that lacks the high-energy photons necessary to create such ions because it has been transmitted through the high-velocity part of the outflow.
That is, in the presence of a multiple gas clouds along a line of sight, the gas cloud further from the radiation source may see an absorbed ``filtered'' SED from the back of the gas cloud that is located closer to the radiation source.
This phenomenon has been investigated previosuly by \citet{leighly18a}, they explored the potential possibilities for the radiation filtering with SDSS~J0850+4451 by creating synthetic spectra using the filtered SEDs.
Both the accelerating and decelerating outflow scenarios with radiation filtering produced features that are not seen in the spectra of SDSS~J0850+4451 and they concluded no support for the radiation shielding of outflowing gas in that object.
\citet{miller18} analyzed the BAL troughs in LBQS~1206+1052 considering the possible ``shading effect'' using photoionization modeling and suggested that the two-phase model was consistent with the data, but was not statistically distinguishable from a one-phase model; that is, the two-phase model was not required for the data.
SDSS~J1352+4239, on the other hand, seems to require an absorption component (zero-velocity component) originating from an absorbed SED to avoid creating the high-excitation ions at zero-velocity and it is not possible to do so with an unabsorbed SED.
The evidence is that we see several moderate to strong low-ionization absorption lines (e.g. \ion{Mg}{2}, \ion{Al}{3}) from the zero-velocity component but the high-ionization lines normally associated with those lines are completely absent from the spectrum.

\begin{figure*}[!t]
\epsscale{1.0}
\begin{center}
\includegraphics[width=5truein]{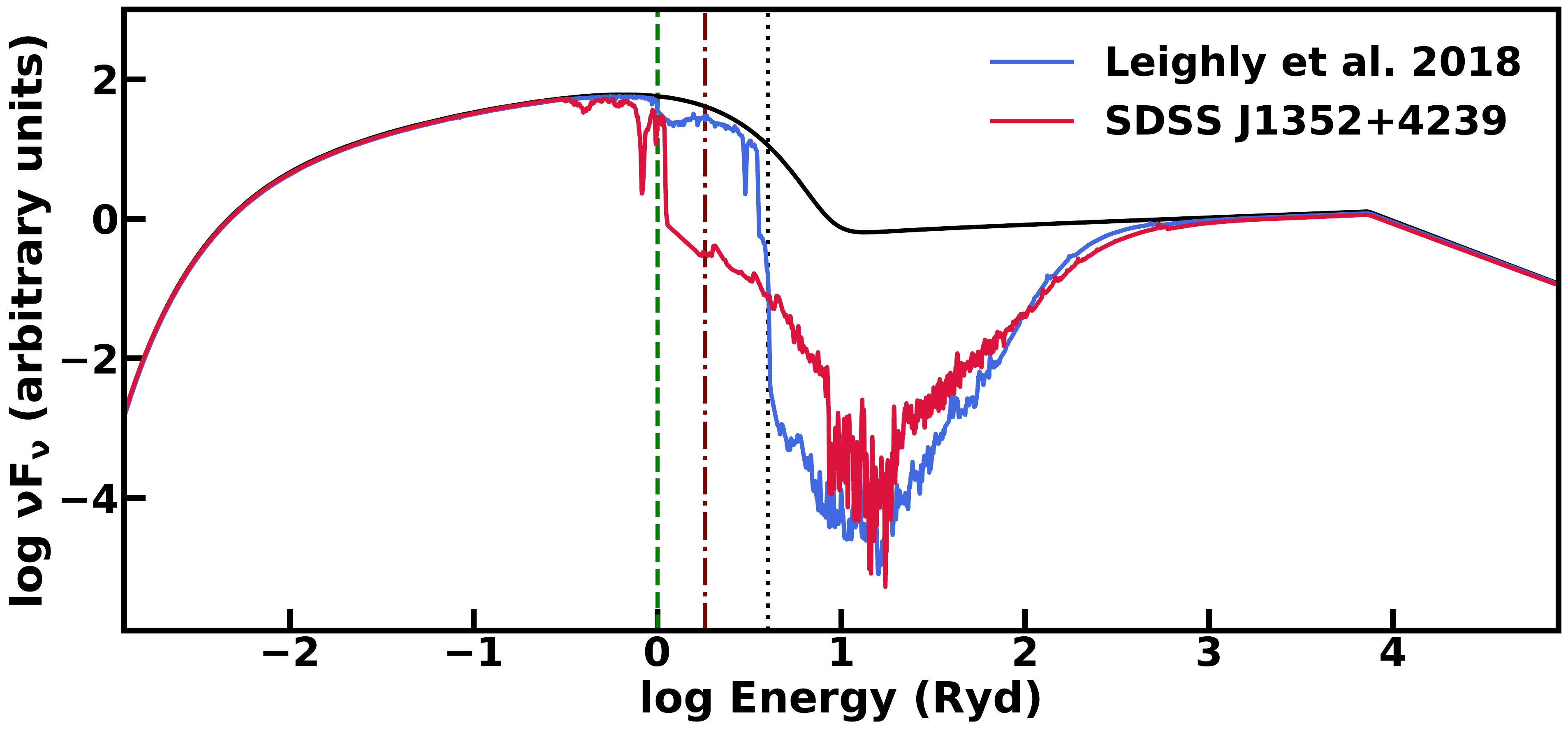}
\caption{The unabsorbed AGN SED is plotted in black and the filtered SED generated from {\it Cloudy} with the physical parameters retrieved from the {\it SimBAL} fit of the blueshifted component is plotted in red.
The filtered SED from SDSS~J0850+4451 \citep{leighly18a} is plotted in blue as a comparison.
The green dashed vertical line, brown dot-dashed vertical line and the black vertical dotted line show the ionizing potentials for \ion{H}{1}, \ion{He}{1}, and \ion{He}{2}, respectively.
SDSS~J1352+4239 shows stronger attenuation in the Lyman continuum, as expected for high column density FeLoBAL, than the LoBAL SDSS~J0850+4451 which has a thinner outflow that does not encompass the hydrogen ionization front.
\label{radshield}}
\end{center}
\end{figure*}

To test the filtering model, we first tried using a modified line list to model the zero-velocity component.
We removed the high-ionization ion transitions (ionization potential $>$ 24.6eV) to approximate such a condition.
The results are not shown, but the success from this approach led to modeling with filtered continuum constructed following \citet{leighly18a} Appendix A.2.
We started with an unabsorbed SED redshifted to match the outflow velocity of the starting bin (highest velocity bin and lowest velocity bin for the decelerating outflow and accelerating outflow, respectively).
Then we created the first transmitted continuum from the starting bin with {\it Cloudy} and used the resulting transmitted continuum to illuminate the next adjacent bin for a subsequent {\it Cloudy} simulation to create the next transmitted continuum.
The final filtered SED for the high-velocity trough was calculated from the transmitted continuum of the final bin.
A more detailed description of the construction of the filtered SED can be found in \citet{leighly18a} Appendix A.2.
We use the filtered SED from the accelerating outflow calculation because we do not find a significant difference between the accelerating and the decelerating outflow scenarios.
Figure~\ref{radshield} shows how the filtered SED differs from the unfiltered AGN SED and how the filtered SED for SDSS~J1352+4239, an FeLoBAL, differs from that of SDSS~J0850+4451, a LoBAL.
A new ionic column density grid was calculated using the filtered SED for the zero-velocity component.

\begin{figure*}[!t]
\epsscale{1.0}
\begin{center}
\includegraphics[width=5.5truein]{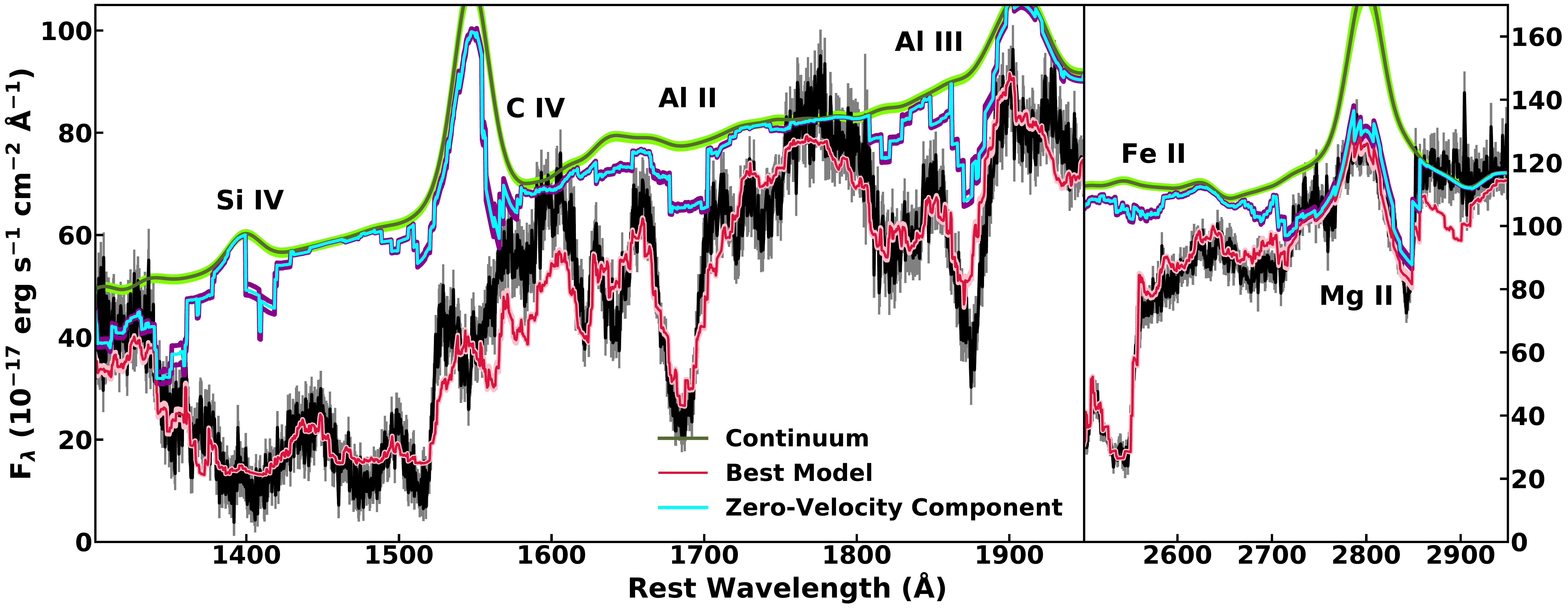}
\caption{The cyan line represents the zero-velocity component model from the filtered grid.
The filtered SED model produces sufficient opacity from the low-ionization ions (\ion{Mg}{2}, \ion{Al}{3} and \ion{Al}{2}) while high-ionization lines (\ion{C}{4} and \ion{Si}{4}) are suppressed.
\label{radshield2}}
\end{center}
\end{figure*}

We fixed the emission and high-velocity trough components from the preliminary best-fitting model and fit only the zero-velocity component with the new column density grid from the filtered continuum.
The physical parameters for the new grid were allowed to vary as fitting parameters.
Figure~\ref{radshield2} shows how the zero-velocity component from the filtered SED produces sufficient low-ionization lines to match the data without overproducing high-ionization lines.
The ionization parameters for the bins ranged between $-2.8$ and $1.8$ with the filtered SED (Table~\ref{param_table}).
The uncertainties associated with the fit parameters and the range of values from the bins for the zero-velocity component were large mainly because the absorption feature is shallow and only a small number of lines are present in the spectrum.

In summary, the absorption feature centered around zero-velocity only showed absorption lines from low-ionization species.
The zero-velocity component from an SED filtered by the high-velocity outflow provided a good fit by producing sufficient opacity for the low-ionization transitions without producing deep high-ionization absorption lines.
The distinction between this result and previous ones looking for evidence for filtering or shading \citep{miller18} is that while the previous efforts found that the data were consistent with filtering, our data show the lack of high-ionization lines that must be the signature of this phenomenon, and therefore require a filtered continuum.

\section{Derived Physical Properties of the Outflow}\label{postpro}
Using {\it SimBAL}, we can measure the physical parameters of the outflow and the uncertainties associated with those values.
We extracted the radius of the outflow using the following relationship:
$$U=\frac{\phi}{nc}=\frac{Q}{4\pi R^2nc},$$
where $\phi$ is the photoionizing flux in, $\rm photons\, s^{-1}\,cm^{-2}$, and $Q$ is the number of photoionizing photons per second emitted from the central engine.
Therefore, with the density and ionization measurements from {\it SimBAL} we can calculate the location of the outflow R.
The value of $Q$ was estimated by scaling the {\it Cloudy} input SED to the quasar spectrum and integrating the scaled SED for energies greater than the hydrogen ionization potential of 13.6 eV.
We estimate log $Q$ = 57.3 - 57.4 [photons s$^{-1}$] when scaled the flux density at 4000 \AA\/ ($F_{4000} = 72.58 \times 10^{-17}\rm \, ergs\, s^{-1}\,cm^{-2}$\AA\/$^{-1}$) and to the near-infrared (rest-optical) photometry, respectively.
\begin{figure*}[!t]
\epsscale{1.0}
\begin{center}
\includegraphics[width=3.5truein]{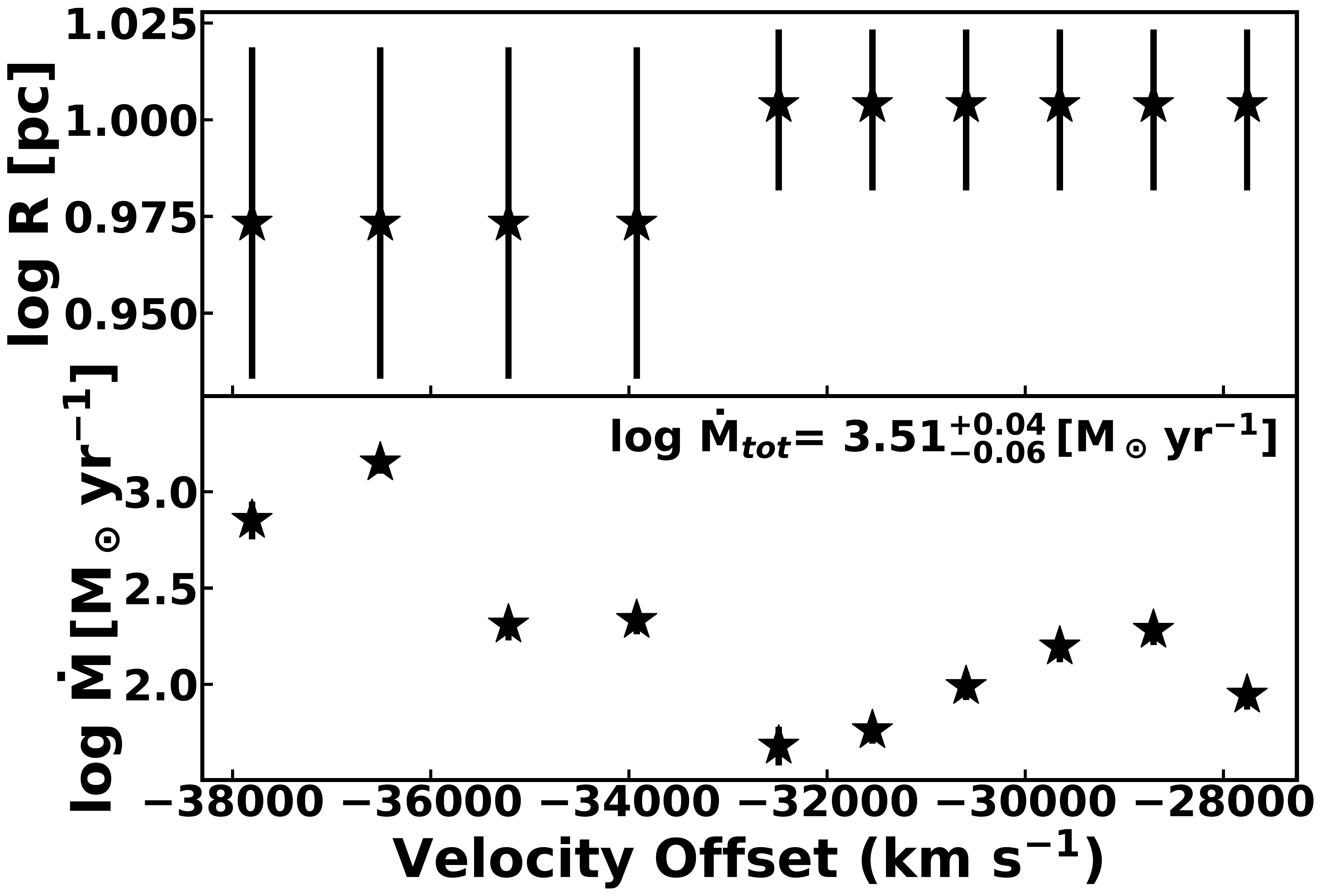}
\caption{The radius and outflow mass estimates for each velocity bin.
The outflowing wind is located $\sim$ 10 pc away from the central engine.
The $\log n$ and $\log U$ values for the bins in the higher and lower velocity groups were constrained to have the same value.
The total outflowing mass of $3200\ (M_\odot\rm\,yr^{-1})$ is noted on the bottom panel.
\label{radiusplot}}  
\end{center}
\end{figure*}
We derived the radius of each bin using the sets of physical parameters constrained by the tophat accordion model (Figure~\ref{radiusplot}).
We found that the location of the outflow is $\sim$ 10 pc away from the center.

Once we know the radius of the outflow, we can further calculate the mass outflow rate of the outflow and the kinetic luminosity associated with it.
We computed the outflow mass using the equation from \citet{dunn10}
$$\dot M=8\pi\mu m_p \Omega R N_H v,$$
where the mean molecular weight is assumed to be $\mu=1.4$, the global covering fraction is given by $\Omega$, and $R$, $N_H$, and $v$ are calculated from the best-fitting parameters from {\it SimBAL}.
We calculate the mass outflow rate for each bin (Figure~\ref{radiusplot}) and sum them to estimate the total mass outflow rate of log $\dot M=3.5\pm0.04\,\rm[M_\odot\,yr^{-1}]$.
The outflowing mass rate of $3210^{+270}_{-290}\ \rm(M_\odot\ yr^{-1})$ is about 18 times the mass accretion rate (\S~\ref{bhmass}).
We use $\Omega=0.2$ based on the fraction of BALQs in optically selected surveys \citep[e.g.,][]{hewett03}, and further discussion of $\Omega$ is below.

Kinetic luminosity is one of the critical physical measures of the outflow strength.
Cosmological simulations require the ratio between the kinetic lumosity and the bolometric luminosity to be 0.5\% to 5\% for effective quasar feedback that could reproduce the observed scaling relations between the host galaxy and the central black hole \citep[e.g.,][]{dimatteo05,hopkins10}.
Using the equation $\dot E_k=\dot Mv^2/2$, we measure the log kinetic luminosity to be 48.1$\pm0.04$ [erg $\rm s^{-1}]$ and $L_{KE}/L_{Bol}$ of $\sim$ 1.
This value of kinetic luminosity is the largest ever found from BAL quasars and sets a new record for the strength of the quasar outflowing wind.
We compare with other large $L_{KE}$ outflows in \S~\ref{energetic_comp}.

In the above mass outflow and kinetic luminosity calculations we adopted the commonly used value of 0.2 for global covering fraction ($\Omega$) following \citet{hewett03} who found 20\% of the optically-selected quasars to have broad absorption lines (once selection effects were accounted for).
The typical values for global covering factor, or the BAL fraction, range from 0.2 to 0.4 depending mainly on the sample selection criteria \citep[e.g.,][]{weymann91,trump06,allen11,dai08,knigge08}.
One explanation for BALs is that they are present in all quasars, covering 20\%$\sim$40\% of the solid angle, and that the fraction of objects with BAL features reflect the amount of sky covered by the quasar outflows in an individual object.
Supporting this view is the fact that (Hi)BALQs have similar broad band spectral energy distribution as the normal quasars \citep[e.g.,][]{gallagher02b,gallagher06,gallagher07}.
However, the above number is derived from HiBALs with \ion{C}{4} lines, and LoBAL fractions can be as low as $\sim$ 1\% in a quasar sample \citep[e.g.,][]{trump06,dai12}.
Assuming this is the case, we would infer the global covering fraction for (Fe)LoBALs to be as low as $\sim$ 0.01.

FeLoBALs can be difficult to identify in the general quasar population due to their lack of strong emission lines and their population fraction might not necessarily reflect the realistic sky coverage of the FeLoBAL wind.
\citet{dunn10} discuss this particular issue in detail and concluded that a selection effect is the reason for the low LoBAL fraction.
They used the value of (Hi)BAL fraction as the global covering fraction for FeLoBAL outflows.
They assert that LoBALs and HiBALs are coming from the physically similar outflowing gas, but we observe LoBAL features from the gas because the light of sight (LOS) happens to pass through the edge of dusty torus.
This not only explains the additional reddening in LoBALs \citep[e.g.,][]{sprayberry92,reichard03} but also the low LoBAL fraction because the LOS needs to be precisely at an angle where it passes through enough torus to produce low ionization lines but not obscure the broad line region.

Finding the true value for BAL fraction or the global covering fraction is difficult and often uncertain.
For example, a large BAL quasar fraction of about $\sim$ 40\% has been found from a luminous infrared selected sample \citep{dai08}.
This value is about double of what \citet{hewett03} found from the optically selected sample but this discrepancy is not very surprising considering BALQs tend to be more frequently reddened than non-BALQs \citep{krawczyk15}.
Therefore, in principle, one can adopt the value of global covering fraction as large as 0.4 for all BALs or as low as 0.01 for FeLoBALs depending on the assumption made to translate the statistical BAL fractions into global covering fractions.

Instead of using a single global covering fraction, we constructed a model to explore the idea that a single outflow exists in the vicinity of the central engine and multiple sightlines observe the outflowing gas as different types of BAL (e.g., HiBAL, LoBAL or FeLoBAL) depending on the viewing angle and the column density the sightline passes through (Fig.~\ref{model_cartoon}).
We estimated the mass outflow rate according to this scenario by gradually lowering the column densities of all the bins by the same small amount while keeping all other parameters fixed to mimic the effect of sightlines passing through less outflowing gas material.
Specifically, we lowered the log $N_H$ $\sbond$ log~$U$ column density parameter and recorded the parameters when the model no longer produced \ion{Fe}{2} absorption lines and transformed to a LoBAL.
We continued lowering the log $N_H$ $\sbond$ log $U$ column density parameter until the \ion{Mg}{2} absorption lines disappeared to create a HiBAL.
From this exercise we were able to estimate log $N_H$ values for different sightlines that can produce different BAL spectral types of the same outflowing cloud reponsible for the trough in SDSS~J1352+4239 ($N_{H HiBAL}$ and $N_{H LoBAL}$).
We then modify the use of single global covering fraction with the following equation
$$
\Omega N_H \Rightarrow \Omega_{HiBAL}N_{H HiBAL}+\Omega_{LoBAL}N_{H LoBAL}+\Omega_{FeLoBAL}N_{H FeLoBAL}.
$$
Using the result from \citet{dai12}, we set $\Omega_{HiBAL}$, $\Omega_{LoBAL}$, and $\Omega_{FeLoBAL}$ to be 0.14, 0.04, and 0.02.
Figure~\ref{model_cartoon} shows the result of our exercise with the changes in the column density noted on the illustration.
We obtain log $L_{KE}$ of $\sim$ 47.6 [erg $\rm s^{-1}$] following the above interpretation.
We conclude that the true value lies between 47.6 (computed using the scenario described here and in Figure~\ref{model_cartoon}) and 48.4 (computed using the maximum value $\Omega$=0.4 from \cite{dai08}).
Applying the same method, we obtain the range of mass outflow rate log $\dot M$ = 3.0 - 3.8 $\rm[M_\odot\,yr^{-1}]$.
We note that the current version of {\it SimBAL} that uses the grids calculated from the version C17 of {\it Cloudy} is only available for the solar metallicity.
A higher metallicity grid would yield a smaller column density and therefore a smaller outflow rate \citep{leighly18a}.

\section{Discussion}\label{disc}
\subsection{A Plausible Geometry of the Outflows}\label{geometry}
In \S~\ref{postpro} we found the radius of the outflow to be approximately 10 pc.
Using the equation $R_{\tau k}=0.47(6\nu L_{\nu}(V))/(10^{46}\rm\,erg\, s^{-1}) $ from \citet{kishimoto07}, derived from near-infrared reverberation monitoring, we estimated the distance to the innermost edge of the torus to be 3.5 pc.
Furthermore, we estimated the dust sublimation radius $R_{sub}\simeq2.0\ \mathrm{pc}$ using the equation $R_{sub}=0.2L_{46}^{1/2} \mathrm{pc}$ from \citet{laor93}.
This indicates that the outflow is located in the vicinity of the dusty torus.

\S~\ref{redcomp} describes the radiation shielding in the zero-velocity component and how this gas must be further from the central engine than the main high-velocity outflow gas.
Considering both the kinematics and the peculiar ionization condition of the absorber, it is possible that the the zero-velocity absorption feature might be arising from an infalling gas cloud.
\citet{hall13} analyzed a sample of objects that show redshifted \ion{C}{4} absorption features and suggested that such absorption signatures can originate from infalling clouds or rotating disk winds.
SDSS~J1352+4239, on the other hand, does not show any redshifted high-ionization lines like the sample \citet{hall13} studied, so it is not possible to use their interpretation of the phenomenon directly.
Also, none of the objects in their sample shows strong blueshifted troughs, therefore it is possible that the physical conditions in SDSS~J1352$+$4239 are very different from their objects.
We speculate that this potential infalling gas could be originating from an earlier ejection episode and we are seeing the signature of the infalling remnant.
\begin{figure*}[!t]
\epsscale{1.0}
\begin{center}
\includegraphics[width=6.5truein]{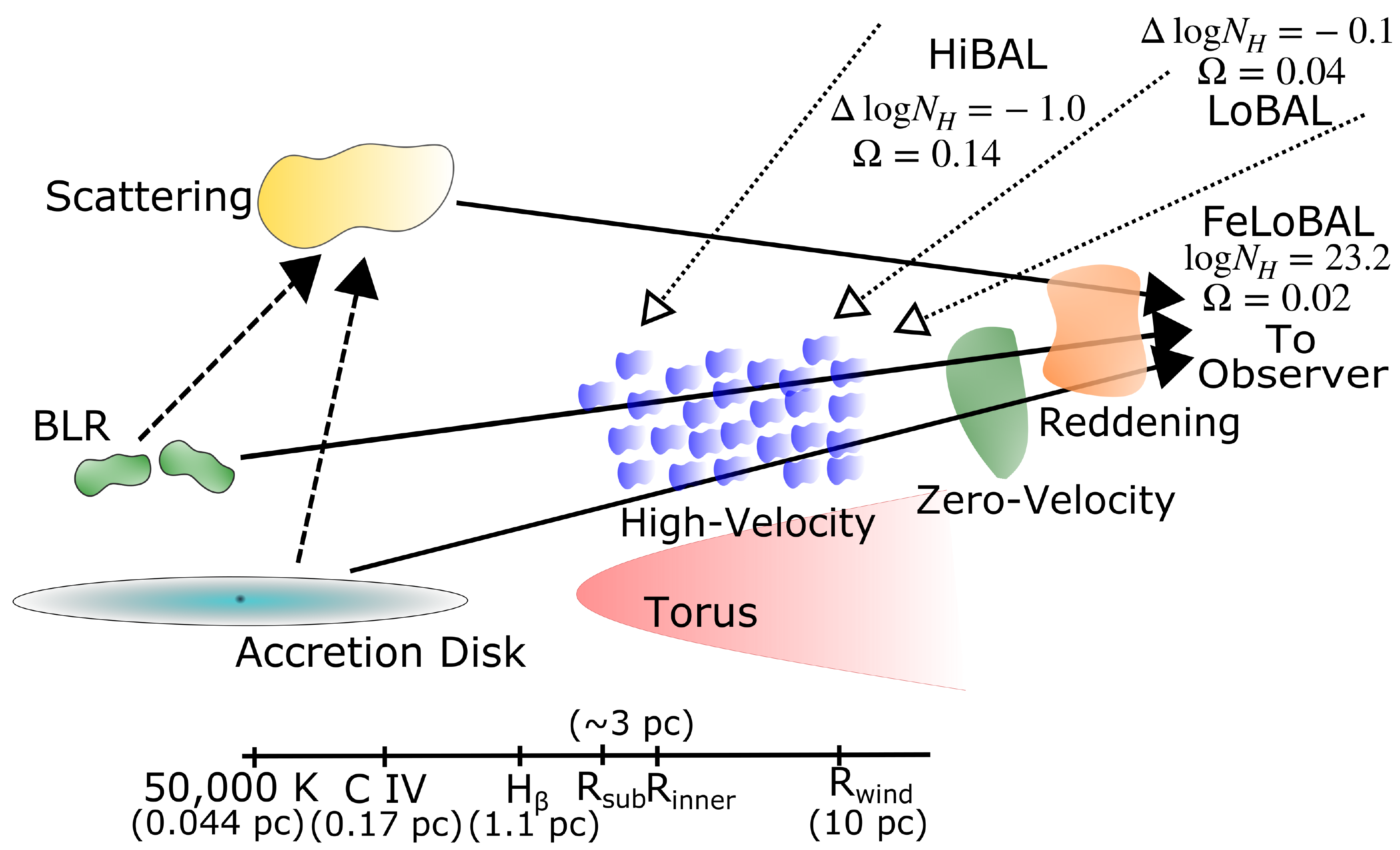}
\caption{The cartoon illustrates how each spectral model component corresponds to different physical AGN components around the central black hole.
The dashed lines represent the photons reaching the scattering medium to create the scattered flux and the solid lines represent the photons reaching the observer.
The dotted lines represent different sightlines for HiBAL, LoBAL and FeLoBAL quasars (\S~\ref{postpro}).
The changes in column density (log $N_H$) required to transform the spectrum from FeLoBAL to the other types and the different global covering fractions ($\Omega$) are labeled on the figure.
The main BAL cloud is located slightly further away from the central engine than the innermost edge of the torus, and the zero-velocity cloud must be located between the main cloud and the reddening source.
The horizontal bar at the bottom of the figure represents the location on the accretion disk where the temperature is about 50,000 K (\S~\ref{accel_mech}), the locations of the \ion{C}{4} and H$\beta$ emitting broad-line regions (\S~\ref{bhmass}), the distances to the torus and the outflowing wind ($R_{inner}$, $R_{sub}$, and $R_{wind}$; \S~\ref{geometry}).
\label{model_cartoon}}
\end{center}
\end{figure*}

Figure~\ref{model_cartoon} shows a physical picture of our spectral model.
From analyzing the best-fitting spectral model, we know the location of the BAL outflow is near the torus.
Both the absorbed spectrum and the scattered flux are reddened, so the dusty reddening source must lie at a larger radius.
The zero-velocity component must be located between the main outflow and the reddening source as the reddening source would transmit too few ionizing photons.
We constrained the ionization parameters for the zero-velocity component to be $\log U<1.8$ and this implies that we can estimate the density of $\log n<5.0\ \rm[cm^{-3}]$ in order for the gas to be located further than the high velocity outflow gas.
We do not have enough information from the spectrum to determine the exact geometry of the scattering cloud.
Potential follow up spectropolarimetry observations may help us gain an insight into the geometry of some of the physical components in SDSS~J1352+4239 we discussed throughout the paper.

\subsection{Acceleration Mechanisms}\label{accel_mech}
We calculated the momentum flux of the outflow from the equation $\dot P=\dot M v$ \citep[e.g.,][]{fg12b}, and we found log $\dot P$ of 38.85$\pm0.04$ [dyne] (38.36 - 39.15 following the global covering fraction interpretation in \S~\ref{postpro} and Fig.~\ref{model_cartoon}) with each individual bin having log $\dot P$ of 37 - 38.5.
Compared to log $L_{Bol}/c$ of 37.5, we find that the ratio between the momentum flux of the outflow and the photon flux is around 20.
The ratio of 20 is far greater than the what is expected of the momentum conserving wind where the maximum momentum flux of the outflow for a single scattering is $L_{Bol}/c$ or momentum flux ratio of $\sim$1 \citep[e.g.,][]{fiore17}.
Two mechanisms have been proposed for objects with large log $\dot P$.
In the energy conserving scenario the outflowing winds get an additional push by the shocks generated from ISM interactions \citep[e.g.,][]{fg12a}.
Such a mechanism can generate a momentum boost and increase the momentum flux ratio between the outflowing gas and radiation by an order of magnitude.
\citet{kp15} discuss various acceleration mechanisms for AGN outflows and compare the size scales of the energy conserving outflows and the momentum conserving outflows.
An energy conserving mechanism mainly explains the $\sim$ kpc size scale outflows where the Compton cooling time-scale becomes greater than the flow time-scale and the full energy of the fast nuclear wind is communicated due to inefficient cooling \citep[e.g.,][]{king11}.
The Compton cooling time for SDSS~J1352+4239 is $t_c\simeq1.16\times10^5R^2_{kpc}\simeq12\rm\ yr$ \citep[Equation (7)]{king11} and we can calculate the flow time $t_{flow}=\frac{R}{v}\simeq330\rm \ yr\ (\it R\sim\rm10\ pc,\it\ v\sim\rm0.1c)$.
It is unlikely for the outflow in SDSS~J1352+4239 to be accelerated via energy conserving mechanism because the cooling is still effective, or the cooling time-scale is smaller than the flow time-scale, $\sim$50 pc and the outflow we found is a compact torus scale outflow (R $\sim$ 10 pc).
The other mechanism involves scattering by dust, which has a larger scattering cross-section than resonance scattering by ions \citep[e.g.,][]{fabian08,fabian18}.
Based on the size scale and the reddening observed in SDSS~J1352+4239, it seems plausible that the outflow is a momentum conserving wind with the additional momentum being harnessed by the dust.
\citet{thompson15} points out that if the effective infrared optical depth is significantly large at the cloud launch point, the outflowing gas can have momentum ratio greater 1 with the momentum conserving mechanism.

We further explored the acceleration mechanism responsible for the high-velocity outflow using force multiplier (FM) analysis.
The FM is defined as the ratio of the total cross-section to the Thompson cross-section.
\begin{figure*}[!t]
\epsscale{1.0}
\begin{center}
\includegraphics[width=4truein]{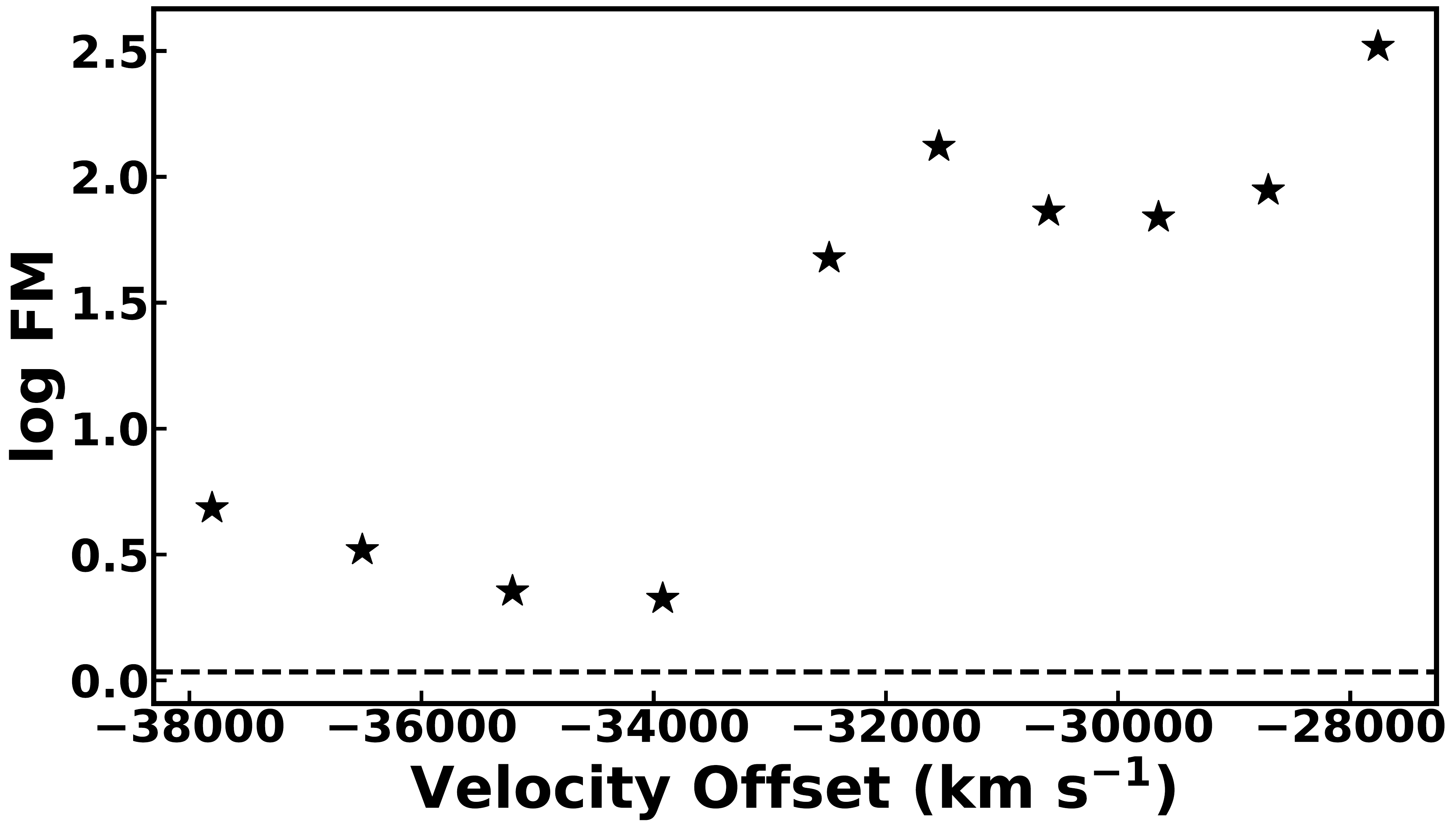}
\caption{The force multiplier (FM) values computed for each bin using {\it Cloudy}.
The horizontal dashed line represents FM = $L_{Edd}$ / $L_{Bol}$ above which the absorber can be radiatively driven.
Because SDSS~J1352+4239 is radiating at near Eddington limit, the FM threshold necessary for the radiative driving is low ($\sim$1) and the FM values for each bin are also rather higher due to lower ionization parameters.
For comparison, see Fig.~17 in \citet{leighly18a} for LoBAL object SDSS~J0850+4451.
\label{force_multi}}
\end{center}
\end{figure*}
We used the best fit parameters from the model and {\it Cloudy} to calculate the force multiplier values for each bin.
Figure~\ref{force_multi} shows the FM values as a function of velocity.
In order for radiative driving of absorbers to occur, FM $\geqslant$ $(L_{Edd}\ /\ L_{Bol})^{-1}$ is a necessary condition \citep[e.g.,][]{netzer13}.
\citet{leighly18a} calculated the FM values for their {\it SimBAL} model of LoBAL object SDSS~J0850+4451 and found that not all tophat bins satisfied the above condition and suggested that alternative driving mechanism might be necessary.
However, SDSS~J0850+4451 is radiating at only 6\% $L_{Edd}$.
SDSS~J1352+4239, on the other hand, is radiating near the Eddington limit (log $(L_{Edd}\ /\ L_{Bol})\ \sim$ 0) therefore even with lower FM values, the absorber can be radiatively driven as all 10 bins have FM values greater than $(L_{Edd}\ /\ L_{Bol})^{-1}$.
This intuitively makes sense since the radiative driving relies on the power of radiation relative to the black hole mass.
The FM values are smaller for the higher velocity bins because they have higher ionization parameter.
Photoionized gas with higher ionization will have fewer ions that can provide UV line opacity and therefore have lower FM.

\begin{figure*}[!t]
\epsscale{1.0}
\begin{center}
\includegraphics[width=4truein]{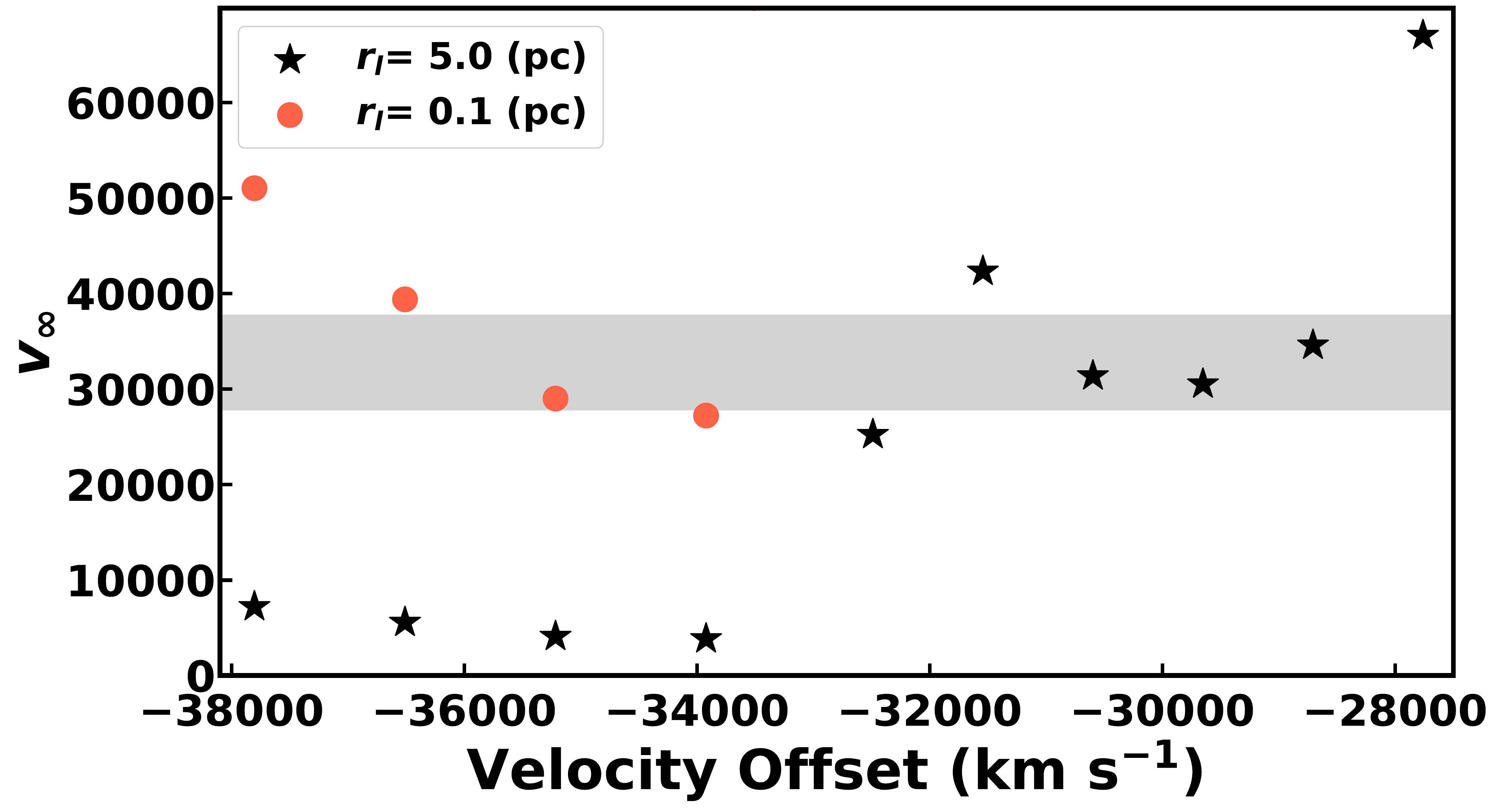}
\caption{The wind terminal velocities for different inner wind radii ($r_l=5.0$ and 0.1 pc in black stars and orange circles, respectively) have been calculated for each bin from the force multiplier (FM) values.
The horizontal gray shaded region shows the actual outflow velocity range observed in SDSS~J1352+4239.
The lower velocity bins can get enough acceleration from large FM values and reach high outflow velocity that we see in the spectra even when launched at a large inner wind radius (5 pc) near the current location of the outflow; however, the higher velocity bins have small FM values (Fig.~\ref{force_multi}) and can only reach high velocity with a smaller launch radius (0.1 pc).
\label{vinf}}
\end{center}
\end{figure*}

FM values alone do not fully explain how the main outflow in SDSS~J1352+4239 was able to reach its high-velocity and large momentum ratio with a large outflow mass.
Therefore we used the equation of motion to further probe how much radiative acceleration can be obtained with the given FM values we found for the main outflow in SDSS~J1352+4239.
We use the equation for acceleration, 
$$v\frac{dv}{dR}\simeq\frac{M(R) \sigma_T L}{4\pi R^2 m_pc} - \frac{GM_{BH}}{R^2}$$
where the first term represents the radiative acceleration with the force multiplier ($M(R)$) and the second term is the force of gravity from the black hole.
Integrating this equation assuming a constant force multiplier value ($FM$) we retrieve the following equation
$$v_{\infty}=32,000R^{-1/2}_{0.1}(6.69\times10^{-3}L_{46}FM-0.008 M_8)^{1/2}\ \mathrm{km\ s^{-1}}$$
where $v_{\infty}$ is the wind terminal velocity, $R_{0.1}$ is the inner wind radius or the launch radius in units of 0.1 pc, $L_{46}$ is the luminosity of the quasar in the units of $10^{46}\ \mathrm{erg\ s^{-1}}$ and $M_8$ is the black hole mass in the units of $10^8\ \mathrm{M_\odot}$.
Figure~\ref{vinf} shows the wind velocities calculated from the above equation.
The wind velocities for the lower velocity bins can reach the observed outflow velocities with the launch radius ($r_l\sim5.0\ \mathrm{pc}$), similar to where we find the outflow ($r\sim10\ \mathrm{pc}$).
But the higher velocity bins require a much smaller launch radius ($r_l<0.1\ \mathrm{pc}$) to match the outflow velocity seen in the spectra.
At such a small radius we expect the gas to be more highly ionized and have smaller FM value, therefore if we compute the integral with FM as a function of radius then the lower velocity bins would need even smaller inner wind radius to be able to reach high outflow velocity.
Note that the above FM values do not include the opacity from the dust.
However with the presence of dust, the total opacity will increase significantly and as a consequence, the gas will be able to obtain extra acceleration.
It will enable the lower velocity bins to potentially reach high velocities even at a larger radius.

Another useful size scale is the location of the UV emission of the accretion disk.
The radiation-driven disk winds are thought to be accelerated by absorption of energetic photons from the UV radiation of the accretion disk \citep[e.g.,][]{proga04}.
The radius at which the disk radiation is mostly in the UV and the location on the accretion disk where the temperature is about 50,000 K is considered the outflow launch radius for such winds \citep[e.g.,][]{giustini19}.
We calculated the location of 50,000 K emission of the accretion disk for SDSS~J1352+4239 to be 0.044 parsec, using the equation $T(R)=(3G M\dot M/8\pi R^3\sigma)^{1/4}$ where $\sigma$ is the Stefan-Boltzmann constant, $M$ is the mass of the black hole and $\dot M$ is the accretion rate.
This value is significantly smaller than the location of the outflow.
Assuming constant outflow velocity of $-30000\ \rm km\ s^{-1}$, it would take about 320 years for the outflow to reach current location of 10 pc if the gas was launched at 50,000 K emission region of the accretion disk.
The value is substantially larger than the rough estimate of the cloud dissipation time \citep[e.g.,][$t\sim\Delta R_{Cloud}/\Delta v\sim10s$ yr for SDSS~J1352+4239]{hamann13}.
Therefore, we suspect the outflow is being radiative driven by both the absorption lines and dust, launched near the torus at a large distance away from the disk.

For example, \citet{czerny17} discuss a failed radiatively accelerated dusty outflow (FRADO) model to understand the motion of the clouds within the broad line region.
Their model is for the broad line region but it is possible that some of the clouds elevated by radiation pressure from the disk or dust would be entrained into the outflow.
And these dusty gas clouds with high opacity can form an outflow that can potentially create BAL troughs.

\subsection{Comparison with Other Known Energetic Quasar Outflows}\label{energetic_comp}
\begin{deluxetable}{lcccccl}
\tablewidth{0pt}
\tabletypesize{\footnotesize}
\tablecaption{Comparison with Other BAL Quasar Outflows\label{energytab}}
\tablehead{
 \colhead{Object} & 
 \colhead{log $L_{Bol}$} & 
 \colhead{log $M_{BH}$} &
 \colhead{$\dot M$} & 
 \colhead{log $L_{KE}$} & 
 \colhead{$\Omega$} & 
 \colhead{Reference}\\ &
 \colhead{[erg $\rm s^{-1}$]}&
 \colhead{[M$_\odot$]}& 
 \colhead{(M$_\odot\rm\,yr^{-1}$)} & 
 \colhead{[erg $\rm s^{-1}$]} & }
\startdata
SDSS~J1106+1939 (LoBAL)&47.2&8.9&390$^{+300}_{-10}$&46.0$^{+0.3}_{-0.1}$&0.08&\citet{borguet13}\\
SDSS~J0831+0354 (LoBAL)&46.9&8.8&410$^{+530}_{-220}$&45.7$^{+0.3}_{-0.4}$&0.08&\citet{chamberlain15}\\
HE~0238-1904 (HiBAL)&47.2&-&69$^{+50}_{-50}$&45.4$^{+0.3}_{-0.6}$&0.5&\citet{arav13}\\
APM~08279+5255 (UFO)&47.45&10.0&11.2&46.9&-&\citet{chartas09,fiore17}\\
SDSS~J1352+4239 (FeLoBAL)&48.0&9.9&1040--6460&47.6--48.4&see \S~\ref{postpro}&This work\\
\enddata
\tablecomments{The mass outflow rate and the kinetic luminosity of the outflow in SDSS~J1352+4239 were estimated using multiple global fractions (\S~\ref{postpro}).}
\end{deluxetable}
We compared our results with other exceptionally energetic outflows in the literature (Table~\ref{energytab}).
\citet{borguet13} found an outflow with log $L_{KE}$ of at least 46 [erg $\rm s^{-1}$] in SDSS~J1106+1939 and it was the most energetic BALQSO outflow ever reported at the time of publication.
SDSS~J0831+0354 was also discovered to have a strong outflow with with log $L_{KE}$ = 45.7 [erg $\rm s^{-1}$] \citep{chamberlain15}.
Since their discovery, several more BAL quasars with comparable energetics have been found.
\citet{fiore17} collected a large sample of AGN outflow data and performed a quantitative analysis on the properties of the outflows.
Some ultra-fast outflow (UFO) objects with absorption lines in the X-ray band have strong winds in their systems due to the high velocity of the outflows.
APM~08279+5255 is a lensed quasar with an X-ray UFO feature that has a near-relativistic outflow with log $L_{KE}$ = 46.85 [erg $\rm s^{-1}$] \citep{chartas09}.
The energy of the outflows we discovered in SDSS~J1352+4239 is greater than even the most energetic UFO outflow known.
Estimating the outflow radius is crucial in estimating the kinetic luminosity of the outflows and it is worth noting that the outflow radius calculation for UFOs are different from the BALQs.
To estimate the radius, the density of the gas needs to be carefully constrained.
For BAL spectra, the density of the gas can be directly constrained by analyzing the density sensitive absorption lines, on the other hand, UFOs and X-ray spectra rely on an indirect method where the density is estimated by interpreting the trough variability \citep[e.g.,][]{risaliti02,hemler19}.
Among the objects listed in Table~\ref{energytab}, SDSS~J1352+4239 is the only FeLoBAL object and the most luminous.
FeLoBAL objects are known to have higher column density relative to the hydrogen ionization front \citep{lucy14} than the other BAL objects and it is possible that in a large FeLoBAL sample we might be able to find more BAL objects with comparable or more energetic outflows \citep{leighly_prep,dabbieri_prep}.

\subsection{How Special is SDSS~J1352+4239?}\label{rarity}
SDSS~J1352+4239 is a very luminous quasar with an energetic outflow and an impressive overlapping trough feature in the rest-UV spectrum.
The quasar luminosity function shows that such luminous quasars are rare objects in the universe with space densities 1$\sim$2 orders of magnitude lower than the less luminous quasars \citep{richards06qlf}.
Moreover, fewer than half of quasars show BAL features (e.g., \citealt{hewett03} ($\sim\,20\%$); \citealt{dai08} ($\sim\,40\%$)) and among the BAL quasars, only a handful of objects show features of very powerful outflows \citep[e.g.,][]{fiore17}.
This means one can find only about $2\sim4$ luminous BALQs that may potentially have strong outflows from a sample of 1000 quasars and a sample of at least tens of thousands quasars is needed to find one luminous quasar with such a high velocity FeLoBAL outflow.
From these statistics, we can infer that SDSS~J1352+4239 is indeed a rare and a special kind of object.

Observational survey programs and the pipelines they use have biases and observational limitations that would result in under-reporting of the BAL quasars with strong outflows or peculiar spectroscopic features (extreme BAL troughs, heavy reddening, and low luminosity and signal-to-noise ratio).
BALQSOs with strong absorption from thick absorbing gas often do not show any strong emission features, making it difficult for survey pipelines to correctly categorize them as quasars.
Strong reddening not only dims the object but it can further make the spectra more difficult to analyze and classify.
More BAL objects similar to SDSS~J1352+4239 may already be in the publicly available archives.

\subsection{Implications for AGN Feedback and Evolution}\label{feedback}
Theoretical model calculations require outflows to have the kinetic luminosities of about 0.5$\sim$5\% of the bolometric luminosity to contribute to AGN feedback and influence the star formation in the host galaxies \citep[e.g.,][]{dimatteo05,scannapieco04,hopkins10}.
The energy in the outflow we discovered in SDSS~J1352+4239 is roughly the same as the quasar bolometric luminosity and we can confidently conclude that the outflow has more than enough energy to influence the star formation in the host galaxy and provide feedback.
The strength of the outflow ($L_{KE}$) is thought to scale with the bolometric luminosity of the quasar \citep[e.g.,][]{costa14,zubovas12}.
SDSS~J1352+4239 has a very high bolometric luminosity, greater than most of the quasars known to have extreme AGN luminosities \citep[e.g.,][WISE/SDSS selected hyper-luminous (WISSH) quasars]{bischetti17}, and the observed energetic outflow (\S~\ref{postpro}, \S~\ref{bhmass}) which seems to support this conjecture.

Some extremely red quasars are also found to have high bolometric luminosities and a fraction of them are known to host strong outflows \citep[e.g.,][]{hamann17,zakamska19}.
\citet{urrutia09} found an anomalously large fraction of BALs (LoBALs) in a sample of red quasars and argues that the LoBAL quasars represent quasars in their early evolutionary stage.
They further suggest the idea that the BAL outflows occur just after the merger events during a ``blow out'' phase which suppresses the star formation in the host galaxy.
Obscured quasars are expected to show a sign of ongoing merger activities and/or a signature of recent star burst episode \citep{sanders88a}; however, the observational evidence shows mixed evidence for merger activities or starbursts \citep[e.g.,][]{violino16,zakamska19}.

\begin{figure*}[!t]
\epsscale{1.0}
\begin{center}
\includegraphics[width=3.5truein]{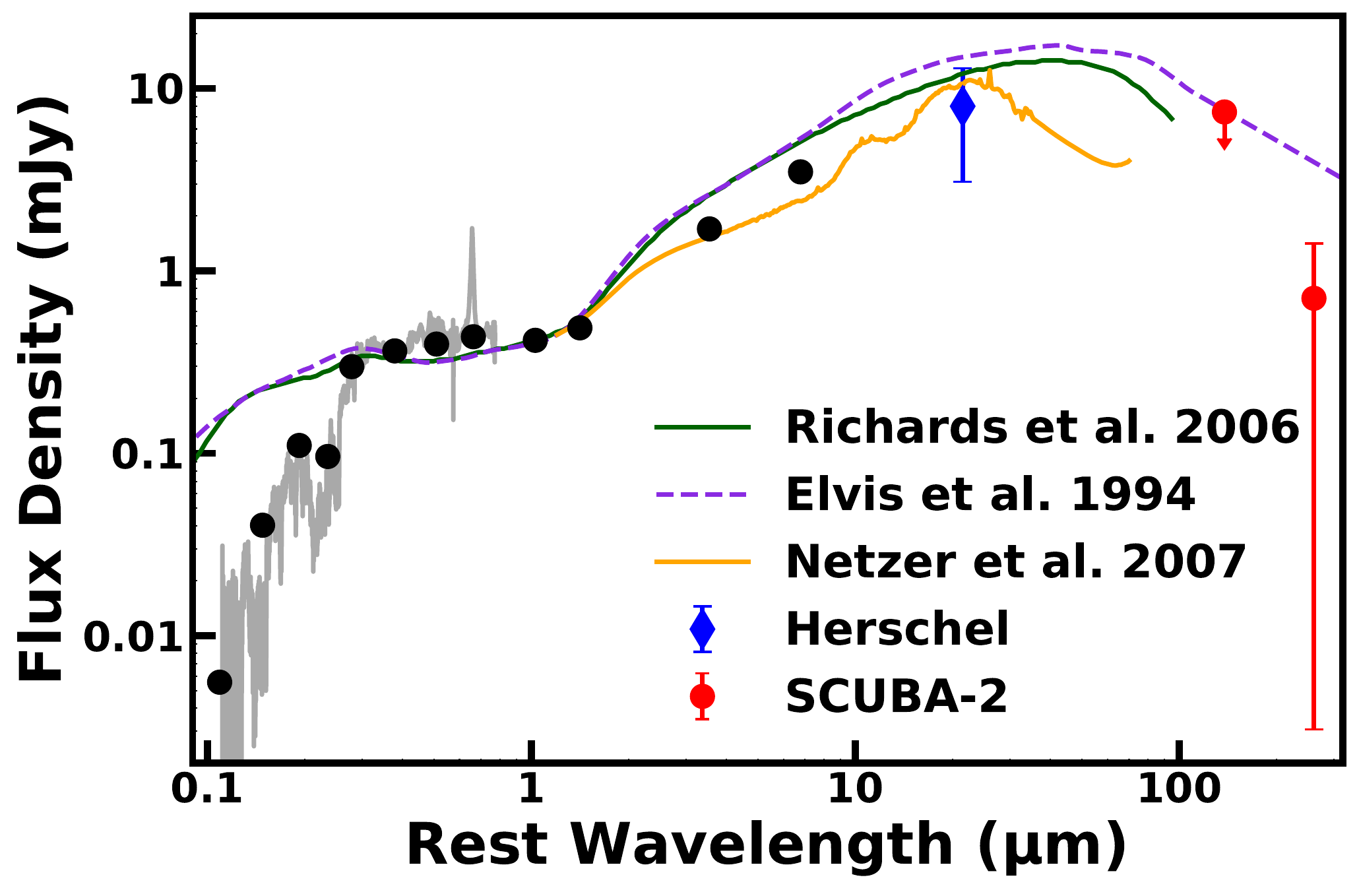}
\caption{The broadband photometry data for SDSS~J1352+4239 is plotted with mean quasar SEDs from \citet{richards06} and \citet{elvis94}.
Both of these SEDs do not account for star formation, so the quasar intrinsic SED from \citet{netzer07} is plotted in orange as well.
Black dots are the photometry data from SDSS, 2MASS and WISE as described in \S~\ref{contsedmod} and shown in Fig.~\ref{fig1}.
The blue dot is the photometry data from Herschel at 70 microns, observed frame.
The red dots are the SCUBA-2 data from \citet{violino16} at 850 microns and 450 microns, observed frame.
The WISE photometry points and Herschel observation of SDSS~J1352+4239 are consistent with the intrinsic quasar SED.
The starburst component would dominate the SED at around 100 microns if there were enhanced star formation in this quasar \citep[e.g.,][]{farrah12}.
We do not see such a far-infrared excess and therefore conclude that there is no strong starburst contribution in SDSS~J1352+4239.
\label{ir_plot}}
\end{center}
\end{figure*}

SDSS~J1352+4239 does not show a signatures of substantial star formation.
\citet{violino16} used the Submillimetre Common-User Bolometer Array 2 (SCUBA-2) to investigate whether FeLoBALs represent an evolutionary step between ultraluminous infrared galaxies (ULIRGs) and unobscured quasars.
They found no evidence for enhanced star formation in FeLoBALs including SDSS~J1352+4239.
SDSS~J1352+4239 was also observed by ESA Herschel Space Observatory \citep{pilbratt10}\footnote{PI: Meisenheimer, ``The Dusty Young Universe: Photometry and Spectroscopy of Quasars at z$>$2''} with PACS \citep{poglitsch10} and SPIRE \citep{griffin10}, and was detected with PACS at 70 microns.
We obtained the PACS data from the Herschel Science Archive\footnote{http://archives.esac.esa.int/hsa/whsa}.
The infrared data are plotted in Figure~\ref{ir_plot} along with composite quasar SEDs from \citet{richards06}, \citet{elvis94} and \citet{netzer07}.
No far-infrared excess is detected.
Therefore the photometry data do not support the need for an extra SED component from a starbust.

\section{Summary}\label{conclusions}

In recent years, several discoveries of powerful AGN outflows have been made \citep[e.g.,][]{borguet13,fiore17,chartas09}.
A number of such discoveries were made from the studies of X-ray observations or emission lines in the optical or mm bands.
UV outflows from BAL quasars have received less attention even though their discovery predates the other channels by decades.
There has not been a well-defined statistical analysis of the BAL absorbers primarily due the complex nature of the BAL spectra.
{\it SimBAL} \citep{leighly18a} enables the first quantitative and systemic studies of UV BAL outflows and their potential for feedback.
With {\it SimBAL}, we were able to analyze the complex absorption features in the overlapping trough quasar spectrum of SDSS~J1352+4239 and discover the most energetic AGN wind discovered to date with log kinetic luminosity of $48.1\pm0.04$ [erg $\rm s^{-1}$].
Our principal results are as follows:

\begin{enumerate}
\item In \S~\ref{correctz}, we used H$\alpha$ to measure the true redshift of $2.2639 \pm 0.0008$, a value about $\Delta z\sim0.25$ larger than the previously reported values for SDSS~J1352+4239.
The true redshift led to the discovery of the extreme velocity of the outflow.

\item The black hole mass calculated from the H$\beta$ line is 8.6 $\times\,10^9\,M_{\odot}$ and $L_{Edd}$ for the given black hole mass is 1.08 $\times\,10^{48}$ [erg $\rm s^{-1}]$ (\S~\ref{bhmass}).
SDSS~J1352+4239 is radiating near the Eddington limit with $\log\ L_{Bol}=48.0\ \rm [erg\ s^{-1}]$ with the mass accretion rate of 176 M$_{\odot}$ per year (\S~\ref{redlong}).

\item In \S~\ref{bestmodel}, we discussed the kinematics and the physical conditions associated with the outflow in SDSS~J1352+4239.
Our model finds the maximum wind velocity of $\sim-38000 \rm \, km\, s^{-1}$ making it the fastest FeLoBAL outflow ever found.
We estimate the total covering-fraction-weighted column density of log $N_H=23.22\pm0.05\,[\rm cm^{-2}]$.

\item In \S~\ref{postpro}, we measured the mass outflow rate of $3210^{+270}_{-290}\ \rm(M_\odot\ yr^{-1})$ with the global covering fraction $\Omega=0.2$.
The mass outflow rate is about 18 times higher than the mass accretion rate.
We found that this outflow has the largest kinetic luminosity ever found with $\log\ L_{KE}=48.1\pm0.04\ \rm [erg\ s^{-1}]$.
For an estimated log $L_{Bol}$ of 48 [erg $\rm s^{-1}]$, we calculate the ratio $L_{KE}/L_{Bol}\sim1$, much greater than the 0.5--5\% thought to be sufficient to contribute to galaxy feedback.

\item We report the first definitive case where the data require a model component generated from a filtered SED, providing a strong support for the radiation shielding in action (\S~\ref{redcomp}).
We conclude that this additional absorber is being irradiated with the AGN SED, but with significant amount of ionizing photons taken out by the fast outflow that is located closer to the central engine.

\item In \S~\ref{geometry}, we found that the outflow is located near the torus.
However, the ratio between the outflow momentum flux and the quasar photon flux is far greater than unity ($\sim$ 20), expected for nuclear/torus scale outflows, suggesting that the extra source of momentum boost is required to explain the dynamics of the outflow we see in SDSS~J1352+4239.
The dust in the environment near torus could potentially serve as the acceleration mechanism (\S~\ref{accel_mech}).

\end{enumerate}

Currently we are analyzing a sample of FeLoBAL objects with {\it SimBAL} \citep{leighly_prep}, and further effort toward creating large sample of quasars with FeLoBAL outflows using machine learning techniques is currently underway \citep{dabbieri_prep}.

\acknowledgements{}
The author thanks Dr.\ Karen Leighly for her constructive feedback and advising and the current {\it SimBAL} group: Dr.\ Donald Terndrup, Collin Dabbieri, Ryan Hazlett and Collin McLeod.
The work is funded by NSF grant AST-1518382 to the University of Oklahoma.

This work is based on observations obtained at the Gemini Observatory, which is operated by the Association of Universities for Research in Astronomy, Inc., under a cooperative agreement with the NSF on behalf of the Gemini partnership: the National Science Foundation (United States), National Science and Engineering Research Council (Canada), CONICYT (Chile), Ministerio de Ciencia, Tecnolog\'{i}a e Innovaci\'{o}n Productiva (Argentina), Minist\'{e}rio da Ci\^{e}ncia, Tecnologia e Inova\c{c}\~{a}o (Brazil), and Korea Astronomy and Space Science Institute (Republic of Korea).
This work is based on observations obtained with the Apache Point Observatory 3.5-meter telescope, which is owned and operated by the Astrophysical Research Consortium.
The computing for this project was partly performed at the OU Supercomputing Center for Education \& Research (OSCER) at the University of Oklahoma (OU).

The authors wish to recognize and acknowledge the very significant cultural role and reverence that the summit of Mauna Kea has always had within the indigenous Hawaiian community. We are most fortunate to have the opportunity to conduct observations from this mountain.

\facility{Gemini:Gillett (GNIRS), ARC: 3.5m (Triplespec), Herschel (PACS)}

\software{emcee \citep{emcee}, Sherpa
 \citep{freeman01}, SimBAL \citep{leighly18a}, Cloudy \citep{ferland17}}

\end{document}